\documentclass[sort&compress,aps,a4paper,superscriptaddress,twocolumn]{revtex4}
\usepackage{amsmath}
\usepackage[dvips]{graphicx}
\usepackage{epsfig}
\usepackage{color}
\usepackage{subfigure}
\usepackage{enumerate}
\usepackage{bm}
\usepackage{amssymb}
\usepackage[breaklinks=true]{hyperref}
\usepackage{relsize} 

\usepackage{soul} 
\usepackage{float} 

\newcommand{\bgeqa}{\begin{eqnarray}}
\newcommand{\edeqa}{\end{eqnarray}}
\newcommand{\vpartb}[1]{{\bf{{#1}}}{}}	
\newcommand{\vpart}[1]{{\textrm{{#1}}}{}}	
\newcommand{\refeq}[1]{Eq.~(\ref{#1})}

\definecolor{darkgreen}{rgb}{0,0.5,0}

\usepackage[]{breakurl}

\hypersetup{
		bookmarksopen=false,
    unicode=false,          
    pdftoolbar=true,        
    pdfmenubar=true,        
    pdffitwindow=true,     
    pdfstartview={FitH},    
    pdfpagemode=UseNone,		
    pdftitle={Kinetic corrections from analytic non-Maxwellian distribution functions in magnetized plasmas.},    
		pdfauthor={O. Izacard},     
    pdfnewwindow=true,      
    colorlinks=true,       
    linkcolor=blue,          
    citecolor=blue,        
    filecolor=magenta,      
    urlcolor=blue           
}

\DeclareMathSizes{6}{14}{10}{6}
\usepackage[margin=0.7in]{geometry}
\begin{document}
\title{\vspace{10mm}\begin{flushleft}\begin{large}\bf Kinetic corrections from analytic non-Maxwellian distribution functions in magnetized plasmas.\end{large}\end{flushleft}}
\author{Olivier Izacard}
\email{izacard@llnl.gov}
\affiliation{Lawrence Livermore National Laboratory, 7000 East Avenue, L-637, Livermore, California 94550, USA.}
\date{March 15, 2016\footnote{Pre-print; Published in \textit{Physics of Plasmas}, \textbf{23} (2016) 082504; \url{dx.doi.org/10.1063/1.4960123}}}

\begin{abstract}
In magnetized plasma physics, almost all developed analytic theories assume a Maxwellian distribution function (MDF) and in some cases small deviations are described using the perturbation theory. The deviations with respect to the Maxwellian equilibrium, called kinetic effects, are required to be taken into account specially for fusion reactor plasmas. Generally, because the perturbation theory is not consistent with observed steady-state non-Maxwellians, these kinetic effects are numerically evaluated by very CPU-expensive codes, avoiding the analytic complexity of velocity phase space integrals. 
We develop here a new method based on analytic non-Maxwellian distribution functions constructed from non-orthogonal basis sets in order to (i) use as few parameters as possible, (ii) increase the efficiency to model numerical and experimental non-Maxwellians, (iii) {help to understand unsolved problems such as diagnostics discrepancies} from the physical interpretation of the parameters, and (iv) obtain analytic corrections due to kinetic effects given by a small number of terms and removing the numerical error of the evaluation of velocity phase space integrals. {This work does not attempt to derive new physical effects even if it could be possible to discover one from the better understandings of some unsolved problems, but here we focus on the analytic prediction of kinetic corrections from analytic non-Maxwellians.} 
As applications, examples of analytic kinetic corrections are shown for the secondary electron emission, the Langmuir probe characteristic curve, and the entropy. This is done by using three analytic representations of the distribution function: the Kappa (KDF), the bi-modal or a new interpreted non-Maxwellian (INMDF) distribution function. The existence of INMDFs is proved by new understandings of the experimental discrepancy of the measured electron temperature between two diagnostics in JET. As main results, it is shown that (i) the empirical formula for the secondary electron emission is not consistent with a MDF due to the presence of super-thermal particles, (ii) the super-thermal particles can replace a diffusion parameter in the Langmuir probe current formula and (iii) the entropy can explicitly decrease in presence of sources only for the introduced INMDF without violating the second law of thermodynamics. Moreover, the first order entropy of an infinite number of super-thermal tails stays the same as the entropy of a MDF. The latter demystifies the Maxwell's demon by statistically describing non-isolated systems.
\end{abstract}\maketitle


The observation of non-equilibrium distribution functions is possible in a large number of areas other than laboratory plasma physics such as in astrophysics plasmas, hydrodynamic fluids and molecular dynamics, atomic physics, condensed matter, chemical reactions and in statistics (see Refs.~\cite{Scudder:1992,
Kuhlen:2010,
Leubner:2008,
Leuthausser:1983,
Riley:1975}). We focus here on laboratory plasma physics with charged particles in a magnetic field relevant to tokamaks and spherical tokamaks for the purpose of energy production by fusion energy. Some phenomena occurring in these devices break the symmetry of the Maxwellian distribution function (MDF) such as radio-frequency wave heating, neutral beam injection, ion orbit loss or simply boundary and external conditions (plasma-surface interaction, external magnetic field configuration, etc) (see Refs.~\cite{Karney:1979,
Jaeger:2006,
Hirshman:1981,
Dawson:1971,
Jassby:1977,
Shaing:1992,
Battaglia:2013,
Dahlin:2015}). Another phenomenon that is highly relevant to future fusion reactors is the self-heating by alpha particles produced by fusion reactions~\cite{Matsuura:2009,Hay:2015}. All these phenomena need to be considered in plasma physics in order to better reproduce the dynamics of current tokamaks and predict the confinement of future fusion reactors because to date, there is no rigorous and efficient theory describing non-Maxwellian distributions (NMDF) at finite collisionality. The ability to describe NMDFs is one of the most important unsolved problem in plasma physics. The number of studies about NMDF has been increasing in the last decades, based on some measurements of distribution function as well as on kinetic numerical simulations. Indeed, the Kappa distribution function (KDF) is usually observed in astrophysics~\cite{Scudder:1992,
Leubner:2008,
Livadiotis_2014_Entropy} where different power laws in velocity phase space appear, but it is not usually observed in laboratory plasma physics. Moreover, bi-modal distribution functions (i.e., the sum of two MDFs) are often used in order to better describe the presence of super-thermal particles~\cite{Popov:2009,Jaworski:2013}, but we argue here that the sum of Maxwellians is valid only in two collisional regimes (i.e., when there is no collision or at infinite collisionality). The interpreted non-Maxwellian distribution function (INMDF) introduced here could be the first self-consistent solution for the description of laboratory non-Maxwellians at finite collisionality. The initial purpose of the INMDF introduced here is to more effectively model already known effects. However, {better understandings coming from} the physical interpretation of the INMDF {could} help {in the future} to {intuitively develop theories describing} new physical phenomena. These three ways of describing non-Maxwellians can be generalized as needed by the readers in order to keep manageable velocity phase space integrals. Indeed, the analytic kinetic corrections obtained here are possible with these three non-Maxwellians and our developed method can be used with the creation of new distribution functions describing the phenomenon under investigation. {In order to clarify to the readers the novelty of this work, we emphasize that in comparison to the literature where a large number of orthogonal basis functions are used for the numerical representation of NMDFs~\cite{Peysson_2008_PoP,
Omotani_2013_PPCF,
Hirvijoki_2015_JPP}, we show the advantage to use non-orthogonal basis functions for the efficient analytic description of NMDFs. This allows at the same time the full analytic computation of the velocity phase space integrals with a small number of simple terms, as detailed in the appendices. Indeed, all required velocity phase space integrals which contain the introduced INMDFs are reduced to some special functions (e.g., the hypergeometric $ _1F_1(a,b,z)$ or the incomplete Gamma $\Gamma(x,z)$ functions). However, because these functions are evaluated at specific values, we obtain for the first time all results as function of a small number of terms including simple polynomials and the exponential and Error functions. This method stays valid for a class of NMDFs obtained from our non-orthogonal basis sets. Our analytic description represents a real asset in plasma physics, leading to the resolution of the efficient unification between kinetic and fluid theories~\cite{Izacard_2016_FischSymposium,Izacard_2016_Paper2}. In summary, this work is unique by its capability to analytically link a specific shape of NMDFs to a small number of fluid quantities observable in experiments.} In order to represent the generality and universality of our work, we focus here on a selection of three relatively different analytic theories which are usually associated with MDFs in the literature:
\begin{itemize}
\item[(i)] As an example, there is a recurrent discrepancy between the transport simulated by fluid codes against its measurement in radiated detached divertor plasmas. The divertor plasma is a crucial component for the fusion energy because it is the main material component in direct contact with the plasma and can highly impact the cost efficiency due to the increase of required maintenance. The difference of transport and plasma profiles between experimental measurements and edge simulations seems to be explained by the presence of NMDFs~\cite{Chankin_2007_NF_47a,
Chankin_2007_NF_47b,
Groth_2013_NF_53}. 
The main issue of these edge simulations is the description of the radiation which is very sensitive to kinetic effects. One of the radiations that directly impact the edge (via the floating potential, the drag force on dust and impurity, or the neutral collision) is the secondary electron emission~\cite{Bruining:1954,
Sternglass_1957_PR,
Dionne:1975,
Meyer-Vernet_1982_AA,
Mayer_1988_PRB,
Bacharis_2010_PoP_17,
Raitses:2011,
Raitses:2015,
Shotorban_2015_PRE}. Recent works focused on the secondary electron emission by using a MDF and obtaining an empirical formula~\cite{Bacharis_2010_PoP_17}. We obtain here the analytic formulas for MDFs and NMDFs.
\item[(ii)] Another example is the discrepancy of the Langmuir probe interpretation~\cite{Arslanbekov,Sudit_1994_JAP,
Stangeby_1995_PPCF,
Batishchev_1997_PoP,
Demidov_2002_RSI,
Tskhahaya_2011_JNM,
Sheehan_2011_PoP,
Godyak_2011_JPD,
Jaworski_2012_FED,
Popov_2012_PSST,
Popov_2014_CPP,
Godyak_2015_JAP} with respect to the Thomson scattering measurements~\cite{Luna_2003_RSI,Beausang_2011_RSI} of the electron temperature which has been associated with the presence of super-thermal particles~\cite{Popov:2009,Jaworski:2013} (i.e., bulk and super-thermal populations, both at different thermodynamic equilibrium). The Langmuir probe is one of the most used diagnostic for low temperature plasmas. Advanced concepts allow the measurement of the electric potential, the electron density and temperature or the radial electric field. All of these measurements are possible by interpreting the signal and by using the assumption that the incident electrons are described by a MDF. Then, the presence of NMDF obviously results in discrepancies of the interpretation of physical quantities from the signal. We show here analytic results for MDFs and NMDFs in order to prepare future modified interpretations of experimental data.
\item[(iii)] Finally, the last example which is analytically based on MDFs is our understanding of the entropy at the thermodynamic equilibrium (i.e., the MDF). The entropy, viewed as the degree of disorder, is well known for a MDF. However, the experimental observations of NMDF steady-states has to be addressed in order to have a more general point of view of the entropy, similarly than in Refs.~\cite{Kranys_1970_ARMA,Livadiotis_2014_Entropy}. The analytic result shown here for an INMDF motivates the generalization of thermodynamics for non-isolated systems (extending Refs.~\cite{Bizarro_2011_PRE,Bizarro_2015_PoP}) by using our INMDFs. Moreover, a significant new result shown here is the explicit decrease of the local entropy for some INMDFs.
\end{itemize}

Our work is drawn as follow. The first section describes the usual analytic representations (i.e., the KDF and the sum of MDF) that are well documented in the literature, and contains the foundations of new distribution functions (i.e., the INMDFs) relevant to some experimentally observed NMDFs. Then, this section focuses on the physical motivations of the introduced INMDF and its possible generalization. In Sec.~\ref{ToC.KinCorrections} the kinetic corrections of the secondary electron emission, the Langmuir probe characteristic curve and the entropy are analytically predicted and shown with as relevant plasma physics parameters as possible. Finally, the conclusion is detailed in Sec.~\ref{ToC.CCL}.

\section{Choice of an analytic representation of non-Maxwellian distribution functions}
\label{ToC.NMDFs}
This section deals with the velocity phase space representation of non-Maxwellian distribution functions (NMDFs) using specific analytic forms instead of the usual numerical discretization. In fact, the numerical approximation is not efficient for the description of some NMDFs because of the requirement to use a high number of terms. For example, it is not efficient to use a numerical approximation of a Maxwellian distribution function (MDF) in comparison to its analytic form which uses only 3 variables (density, fluid velocity and temperature). Indeed, this work argues the advantage of using a mesh-free representation of the distribution function in the velocity phase space by using fluid parameters (called here the hidden variables) instead of the discretization in the velocity phase space. We adopt the notation of ${\rm v}$ for the one dimension velocity phase space coordinate, $\vpartb{x}$ for the three dimension position and $t$ for the time. The generalization to three dimensional velocity phase space is ongoing for isotropic distribution functions but some complexities appear for anisotropic ones~\cite{Izacard_2016_Paper2}. Further work will include anisotropic NMDFs. For simplicity of notations the division of the temperature by the mass is omitted in the distribution function. \\
In this section we show notations for existing distributions (the KDF and the sum of two MDFs) and we introduce a new class of NMDFs (i.e., the INMDF) which can be viewed as a non-orthogonal generalization of the sum of Maxwellians and Hermite polynomials.

\subsection{The Kappa distributions}
The KDFs are commonly used in astrophysics~\cite{Scudder:1992,Leubner:2008} to describe different tail power laws than the Maxwellian. The definition of the KDF $f_{\kappa}(\vpartb{x},{\rm v},t)$ is
\bgeqa
\label{eq.nM.fKappa}
\displaystyle
f_{\kappa} &=& \Delta_{\kappa}  \left(1 + \frac{{\rm v}^2}{W_{\kappa}}\right)^{-(\kappa+1)},
\edeqa
where
\bgeqa
\displaystyle
\Delta_{\kappa} &=& \frac{n}{\sqrt{\pi W_{\kappa} }} \left( \frac{\Gamma (\kappa+1)}{\Gamma \left(\kappa-\frac{1}{2}\right)} \right)^{1/3}, \\
\displaystyle
W_{\kappa} &=& (2 \kappa-3) T,
\edeqa
and $\Gamma(z)$ is the Euler Gamma function. The fluid quantities, which are evolving in time and are functions of the position, are $n(\vpartb{x},t)$, $W_{\kappa}(\vpartb{x},t)$ (or $T(\vpartb{x},t)$) and $\kappa(\vpartb{x},t)$. For a shifted distribution function the fluid velocity $v(\vpartb{x},t)$ appears by the change of variable ${\rm v} \rightarrow {\rm v}-v$ in Eqs.~(\ref{eq.nM.fKappa}), (\ref{eq.nM.KqkappaT}) and~(\ref{eq.nM.KqkappaTa}). The KDF recovers the MDF for $\kappa \rightarrow \infty$. Many properties of the KDF have been published in the literature and are not reproduced here. Integrals of KDFs over the velocity phase space introduce functions detailed in App.~\ref{ToC.App.KqkappaTa}
\bgeqa
\label{eq.nM.KqkappaT}
\displaystyle
K_{q}(\kappa ,T) &=& \int_{-\infty}^{\infty} {\rm v}^{q} \left( 1 + \frac{{\rm v}^2}{W_{\kappa}} \right)^{-(\kappa +1)} d{\rm v}, \\
\label{eq.nM.KqkappaTa}
\displaystyle
K_{q}(\kappa ,T,a) &=& \int_{a}^{\infty} {\rm v}^{q} \left( 1 + \frac{{\rm v}^2}{W_{\kappa}} \right)^{-(\kappa +1)} d{\rm v}.
\edeqa
By using these integrals, it is trivial to obtain all fluid moments of any shifted KDF centered at the fluid velocity $v$ as function of a finite sum of the function $K_q(\kappa,T)$. The KDF is successfully used to describe different power laws in velocity phase space. However, because an over population of fast particles localized around a specific velocity is often observed, the sum of two Maxwellian can be a better approximation for these cases.

\subsection{The sum of Maxwellians}
The approximation of localized super-thermal tails is commonly used~\cite{Popov:2009,Jaworski:2013,Jaworski_2015_Private}. It is usually called the bi-modal distribution function where two different temperatures dominate as function of the velocity coordinate ${\rm v}$. The sum of two shifted MDFs $f_{2M}(\vpartb{x},{\rm v},t)$ reads 
\bgeqa
\label{eq.nM.f2Maxw}
\displaystyle
f_{2M} &=& \Delta \exp\left( -\frac{1}{2 T} {\rm v}^2 + \frac{v}{T} {\rm v} \right) \nonumber\\&+& \Delta_f \exp\left( -\frac{1}{2 T_f} {\rm v}^2 + \frac{v_f}{T_f} {\rm v} \right),
\edeqa
with
\bgeqa
\label{eq.nM.f2Maxw.del}
\displaystyle
\Delta &=& \frac{n}{\sqrt{2\pi T}} \exp\left( -\frac{v^2}{2 T} \right), \\
\label{eq.nM.f2Maxw.delf}
\displaystyle
\Delta_f &=& \frac{n_f}{\sqrt{2\pi T_f}} \exp\left( -\frac{v_f^2}{2 T_f} \right),
\edeqa
where the density, fluid velocity and temperature are respectively for the bulk and the super-thermal (i.e., fast particles) populations $n(\vpartb{x},t)$, $v(\vpartb{x},t)$, $T(\vpartb{x},t)$ and $n_f(\vpartb{x},t)$, $v_f(\vpartb{x},t)$, $T_f(\vpartb{x},t)$.
The velocity integrals are directly given by the function $J_k(a,b)$ detailed in App.~\ref{ToC.App.Jkab} and defined by
\bgeqa
\label{eq.Gen.Jkab}
\displaystyle
J_k(a,b) &=& \int_{-\infty}^{\infty} {\rm v}^k \exp\left(-a {\rm v}^2 + b {\rm v}\right) d{\rm v}.
\edeqa
Another useful function is $J_k(a,b,c)$ detailed in App.~\ref{ToC.App.Jkab0} and App.~\ref{ToC.App.Jkabc} and defined by
\bgeqa
\label{eq.Gen.Jkabc}
\displaystyle
J_k(a,b,c) &=& \int_{c}^{\infty} {\rm v}^k \exp\left(-a {\rm v}^2 + b {\rm v}\right) d{\rm v}.
\edeqa
The generalization with a sum of more than two MDFs is possible with the Radial Gaussian Basis Function (RGBF) which has successfully been used for artificial neural networks. Many properties of the bi-modal distribution function and the RGBF have been published in the literature and are not reproduced here. \\
This representation is accurate for example when there is an external production of fast particles and when there is no interaction between the bulk plasma (of collisionality $\nu_{\rm th-th}$) and these fast particles (of collisionality $\nu_{\rm th-f}$). This means that the RGBF is valid only at two collisional regimes: the limit of no collision between thermalized and fast particles (i.e., $\nu_{\rm th-f}=0$) or the limit of infinite collisionality (i.e., $\nu_{\rm th-th}/\nu_{\rm th-f} \rightarrow \infty$). In other words, the RGBF is not consistent with interactions between the bulk plasma and the fast particles at finite collisionality because each of them become non-Maxwellian. The novel representation introduced below is a possible solution of this inconsistency.

\subsection{The interpreted non-Maxwellian}
\label{ToC.NMDFs.INMDF}
\begin{figure}[!ht]
\centering
\begin{subfigure}[]{}
\centering
\includegraphics[height=30mm,natwidth=585,natheight=441]{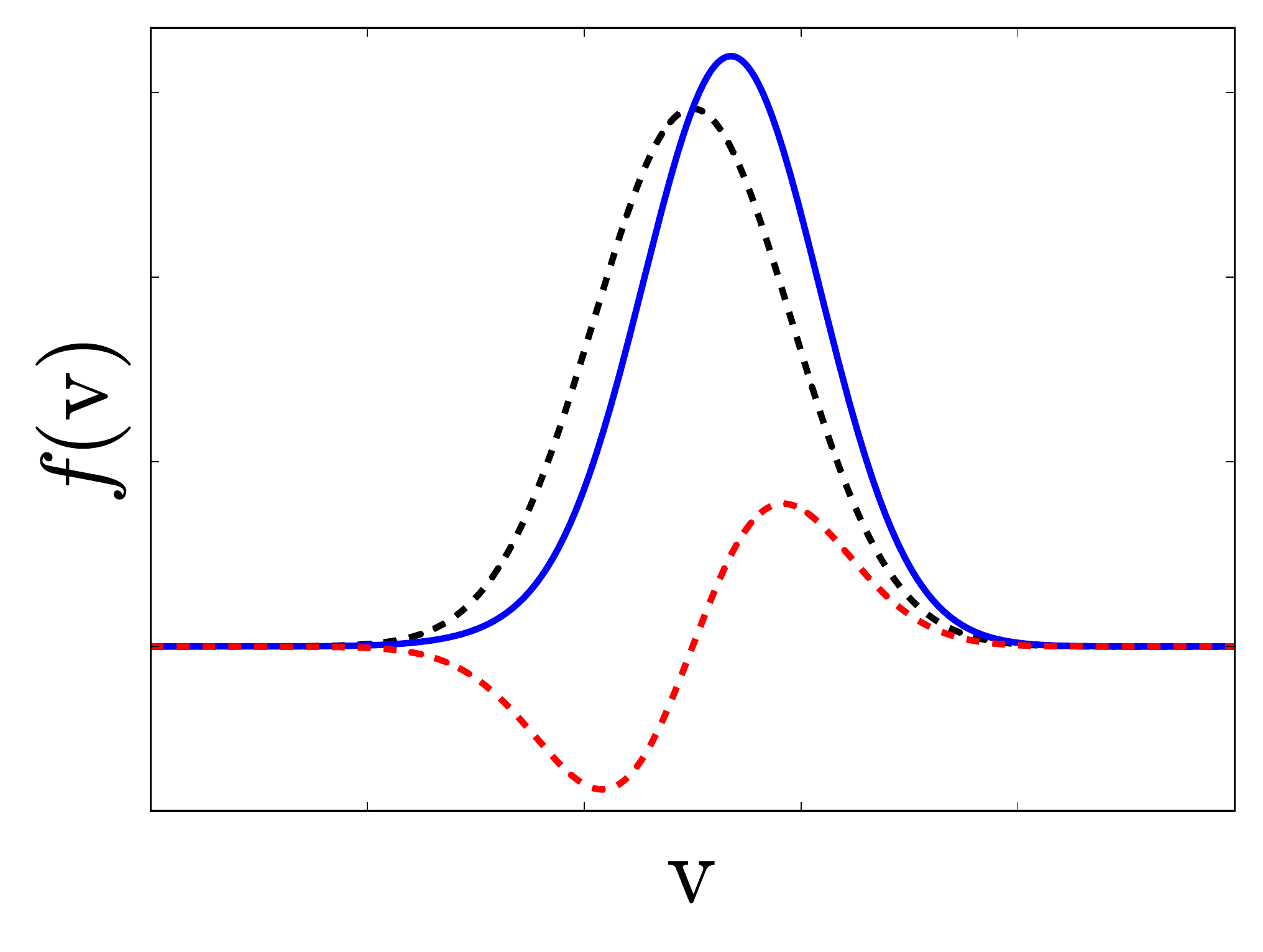}
\label{fig.Izacardian_examples1.1}
\end{subfigure}%
\begin{subfigure}[]{}
\centering
\includegraphics[height=30mm,natwidth=585,natheight=441]{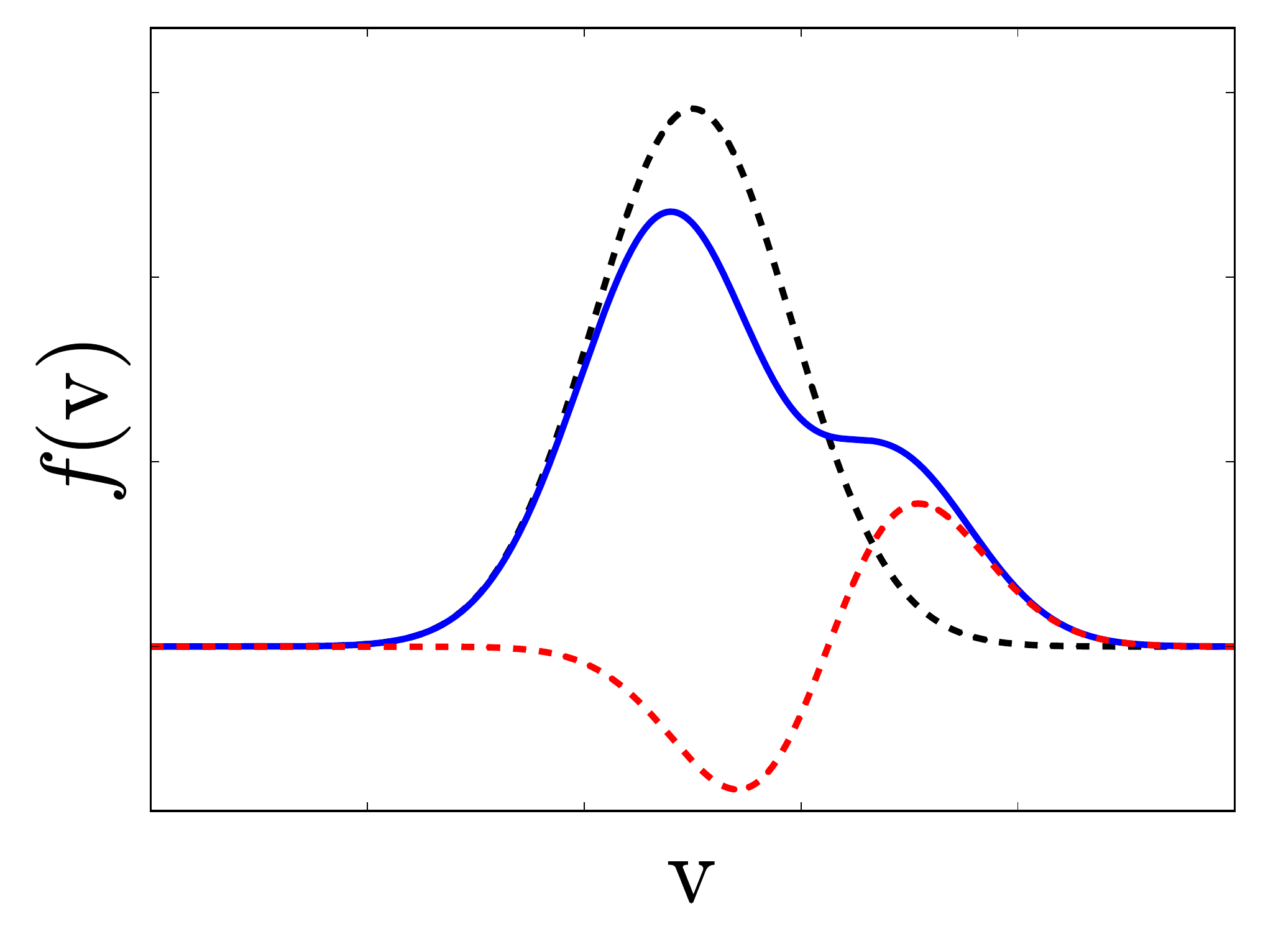}
\label{fig.Izacardian_examples1.2}
\end{subfigure}%
\caption{Schema of the Maxwellian (dashed black curve) and the INMDF (solid blue curve) distribution functions and the difference (dashed red curve). These curves are obtained with the density $n=10^{19}$ m$^{-3}$, the temperature $T=10^4$ eV, the velocity space ${\rm v} \in \left[ 0 , 10^5 \right]$, the number of points in the velocity space $N_{\rm v}=1001$, the fluid velocity $v = ({\rm v}_{max}-{\rm v}_{min})/2$, the kinetic flux $\Gamma=0.1n\sqrt{T}$, the central flow (a) $c=0$ (b) $c=1.25 v$ and the width of the heat spread $W=T/2$. The physical interpretation of these three additional variables is given below.}
\label{fig.Izacardian_examples1}
\end{figure}

The definition of the INMDF $f_I(\vpartb{x},\vpartb{v},t)$ was introduced for the first time in one dimension by Refs.~\cite{Izacard_2013_GA_Talk,Izacard_2013_Emails_Candy,Izacard_2016_FischSymposium}
\bgeqa
\label{eq.nM.fINMDF}
\displaystyle
f_I &=& f_0 + \delta f,
\edeqa
with
\bgeqa
\label{eq.nM.fINMDF.f0}
\displaystyle
f_0 &=&  \Delta \exp\left( -\frac{1}{2 T} {\rm v} + \frac{v}{T} {\rm v} \right), \\
\displaystyle
\label{eq.nM.fINMDF.delf}
\delta f &=& \Delta_I (\vpart{v}-c) \exp\left( -\frac{1}{2 W} {\rm v} + \frac{c}{W} {\rm v} \right),
\edeqa
and with $\Delta$ given by \refeq{eq.nM.f2Maxw.del} and 
\bgeqa
\displaystyle
\Delta_I &=& \frac{\Gamma}{\sqrt{2\pi W^3}} \exp\left(-\frac{c^2}{2W}\right),
\edeqa
where the fluid hidden variables $\{n,v,T,\Gamma,c,W\}$ are function of $({\bf x},t)$. The first part $f_0$ is a MDF and due to the second part $\delta f$, the INMDF $f_I$ cannot be described with a finite number of terms using any of the existing analytic representations. The second part is related to a physical interpretation and could be valid for all collisionality regimes because $\delta f$ represents an enhancement of the energy of a population of particles. This new formula represents a real asset for the analytic modeling of NMDFs. More details of the difference between the INMDF and all existing analytic NMDFs are given in Sec.~\ref{ToC.NMDFs.Gen}. The moments $M_k = \int f {\rm v}^k d{\rm v}$ of the non-Maxwellian $f=f_I$ are
\bgeqa
\label{eq.fI.M0}
\displaystyle
M_0 &=& n, \\
\displaystyle
M_1 &=& n v + \Gamma, \\
\displaystyle
M_2 &=& n \left( T + v^2 \right) + 2 \Gamma c,  \\
\displaystyle
M_3 &=& n v \left( 3 T + v^2 \right) + 3 \Gamma \left( W + c^2 \right), \\
\displaystyle
M_4 &=& n \left( 3 T^2 + 6 T v^2 + v^4 \right) + 4 \Gamma c \left( 3 W + c^2 \right), \\
\displaystyle
M_5 &=& n v \left(15 T^2 + 10 T v^2 + v^4 \right) \nonumber\\&&+ 5 \Gamma \left( 3 W^2 + 6 W c^2 + c^4 \right), \\
\label{eq.fI.M6}
\displaystyle
M_6 &=& n \left(15 T^3 + 45 T^2 v^2 + 15 T v^4 + v^6 \right) \nonumber\\&&+ 6 \Gamma c \left( 15 W^2 + 10 W c^2 + c^4 \right),
\edeqa
and generally given as function of $J_k(a,b)$. The usual moments $P_k = 1/M_0 \int f (\vpart{v}-M_1/M_0)^k d\vpart{v}$ for $k \geq 2$ of the non-Maxwellian $f_I$ are also function of the hidden variables. We do not write these moments since it is not used here, but if needed, readers can easily obtain them from the given $M_k$ moments. 
Nevertheless, we remark as expected for a MDF that when $\Gamma = 0$ the odd moments $P_{2k+1}$ are equal to 0 and the even moments are $P_{2k} = (2k-1)!!\ T^k$. 
The coefficient $\Gamma$ has the dimension of a momentum (a density times a velocity) then $\Gamma/n$ has the dimension of a velocity, $c$ of a velocity, and $W$ of a temperature (square of a velocity, including the division of the temperature by the mass which is omitted by simplicity of notations). \\
We remark that, by using the method of hidden variables, there is no reason to resolve the inversion of the Eqs.~(\ref{eq.fI.M0})-(\ref{eq.fI.M6}) in order to extract the hidden variables as function of the fluid moments. To  date, the inversion of these formulas to extract the hidden variables $\{n,v,T\}$ as function of the moments $\{M_0,M_1,M_2\}=\{n,nv,nT+nv^2\}$ is possible only for a MDF and it seems too constraining to use this criteria for the description of non-Maxwellian steady-state distribution functions observed experimentally. \\
Fig.~\ref{fig.Izacardian_examples1}.(a) represents an INMDF (solid blue curve) with a correction (dashed red curve) centered at the mean velocity with respect to the MDF (dashed black curve). This example corresponds to an asymmetric distribution function observed at least in presence of ion orbit losses~\cite{Battaglia:2013} and will be investigated in future work. Fig.~\ref{fig.Izacardian_examples1}.(b) represents the same correction $\delta f$ but centered at a higher velocity coordinate (i.e., higher value of $c$). 
The tail at high energy (centered at $c>v$) is unstable when the distribution function increases and is not monotonous. In another words, the distribution function is unstable when its first derivative is positive around ${\rm v}=c$. Indeed, the stability criteria of the INMDF $f_I=f_0+\delta f$ is obtained when the derivative of the Maxwellian part $f_0$ is bigger than the derivative of the additional part $\delta f$ (i.e. $\partial f_I/\partial {\rm v}<0$). We found the stability criteria
\bgeqa
\displaystyle
&& \left( 1 - \frac{({\rm v}-c)^2}{W} \right) \frac{\Gamma}{\left(2\pi W^3 \right)^{1/2}} \exp\left( -\frac{\left({\rm v}-c\right)^2}{2 W} \right) \nonumber\\
\displaystyle
&&\quad < \frac{{\rm v}-v}{T} \frac{n}{\left(2\pi T \right)^{1/2}} \exp\left( -\frac{\left({\rm v}-v\right)^2}{2 T} \right),
\edeqa
then, for $W < T$ the highest value at ${\rm v}=c$ gives for $c>v$ the stability criteria
\bgeqa
\label{eq.nM.Izacardian.StabilityCriteria}
\displaystyle
\frac{\Gamma}{n (c-v)} \left(\frac{T}{W}\right)^{3/2} \exp\left( \frac{\left(c-v\right)^2}{2 T} \right) < 1.
\edeqa
For $W > T$, similar stability criterion can be obtained for a range of velocity but are not detailed here. 
Moreover, the constraint to consider a strictly positive distribution function is given by the relation $f_I(\vpart{v})>0$ for all ${\rm v}$, particularly when, for $\vpart{v}<c$, the additional non-Maxwellian part $\delta f$ is minimum (for $\Gamma>0$). The minimum value of the additional non-Maxwellian part $\delta f$ is obtained at
\bgeqa
\displaystyle
\frac{\partial \delta f}{\partial {\rm v}} = 0 &\Leftrightarrow & \frac{({\rm v}-c)^2}{W} = 1, \nonumber\\
\displaystyle
&\Leftrightarrow & {\rm v} = c - \sqrt{W}.
\edeqa
Then the constraint of a strictly positive distribution function is given by $f_I\left({\rm v}=c - \sqrt{W}\right) > 0$, so
\bgeqa
\displaystyle
\frac{\Gamma \sqrt{T}}{n W} \exp\left( \frac{\left(c-v-\sqrt{W}\right)^2}{2 T} - \frac{1}{2} \right) < 1.
\edeqa
We remark that for $\Gamma<0$, similar relations can be obtained using the minimum value of $\delta f$ obtained for ${\rm v}>c$ (i.e., at ${\rm v} = c+\sqrt{W}$). \\
Before investigating corrections of some existing theories due to non-thermal population of particles, we have to understand how the previously given formulas could accurately describe deviations from MDFs. For that, the physical interpretation of the INMDF is given here. 
\begin{figure}[!htbp]
\centering
\includegraphics[height=65mm,natwidth=585,natheight=441]{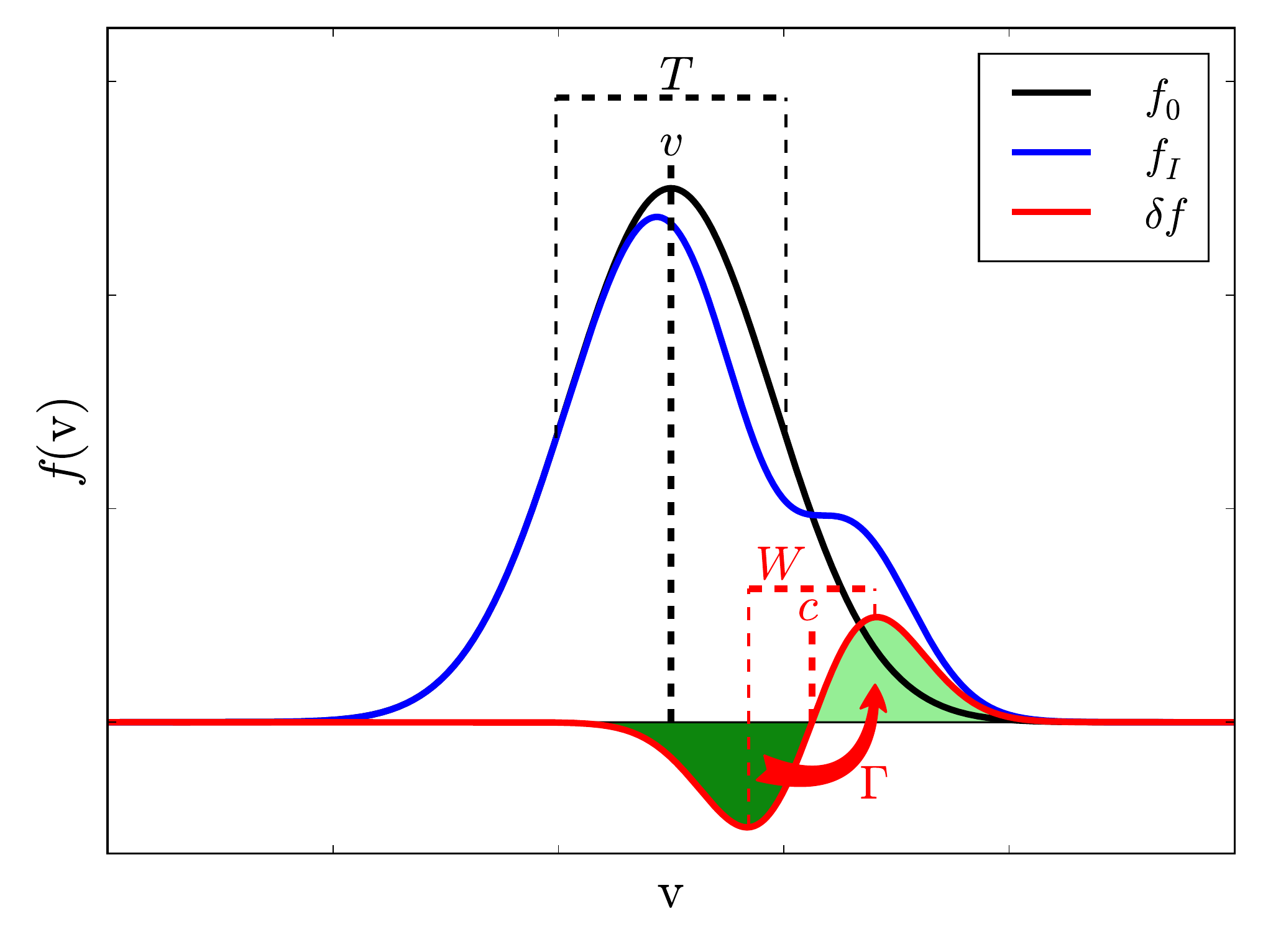}
\caption{Schema of the physical interpretation of the INMDF. The Maxwellian $f_0$ (black curve), the INMDF $f_I$ (blue curve) and the difference $\delta f = f_I-f_0$ (red curve) are shown. These curves are obtained with the density $n=10^{19}$ m$^{-3}$, the temperature $T=10^4$ eV, the velocity space ${\rm v} \in \left[ 0 , 10^5 \right]$, the number of points in the velocity space $N_{\rm v}=1001$, the fluid velocity $v = ({\rm v}_{max}-{\rm v}_{min})/2$, the kinetic flux $\Gamma=0.1n\sqrt{T}$, the central flow $c=v+1.25\sqrt{T}$ and the width of the heat spread $W=T/2$.}
\label{fig.nM.interpretation}
\end{figure}
The fluid quantity $\Gamma({\rm x},t)$ is called here the kinetic flux, $c({\rm x},t)$ the central flow and $W({\rm x},t)$ the width of the heat spread. Similar to the graphical interpretation of $n$, $v$ and $T$ for the MDF, the graphical interpretation of the non-Maxwellian part $\Gamma$, $c$ and $W$ is shown in Fig.~\ref{fig.nM.interpretation}. \\
Fig.~\ref{fig.nM.interpretation} represents the displacement of a population of particles ($\sim 5\%$) from the dark-green area to the light-green area. The hidden variable $\Gamma$ is called the kinetic flux since it happens in the velocity phase space and has a dimension of a particle flux (i.e., a density times a velocity). The central flow $c$ is the velocity where the INMDF equals the MDF. Finally, the width of the heat spread $W$ characterizes the width in velocity phase space of the super-thermal population modified by an external source of energy (e.g., the current drive). In the following figures, the coefficients $(q,r,s)$ are used to parameterize the super-thermal tail such that $\Gamma = 2 q n \sqrt{T}/100$ where $q$ represents approximately the percentage of super-thermal particles over the total number of particles, $r$ represents the position of the central flow by the relation $c=v+r\sqrt{T}$, and $s$ represents the ratio of the width of the heat spread over the temperature $W=s^2 T$ (i.e., $s = \sqrt{W/T}$ is the ratio of the widths).

\subsection{Generalization with other INMDFs}
\label{ToC.NMDFs.Gen}
\begin{figure}[!ht]
\centering
\begin{subfigure}[]{}
\centering
\includegraphics[height=30mm,natwidth=585,natheight=441]{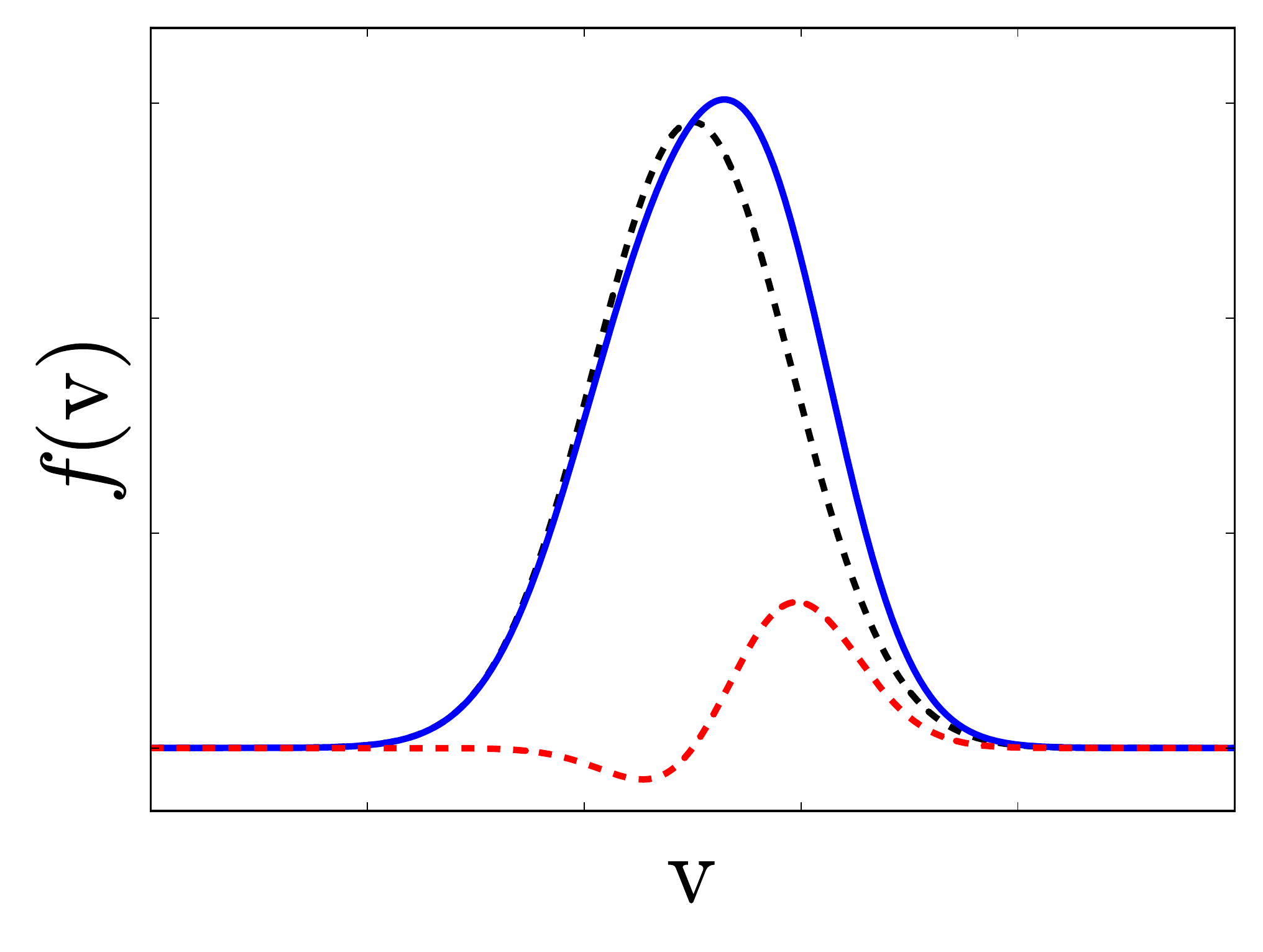}
\label{fig.fnM.Generalization.AsymTail.1}
\end{subfigure}%
\begin{subfigure}[]{}
\centering
\includegraphics[height=30mm,natwidth=585,natheight=441]{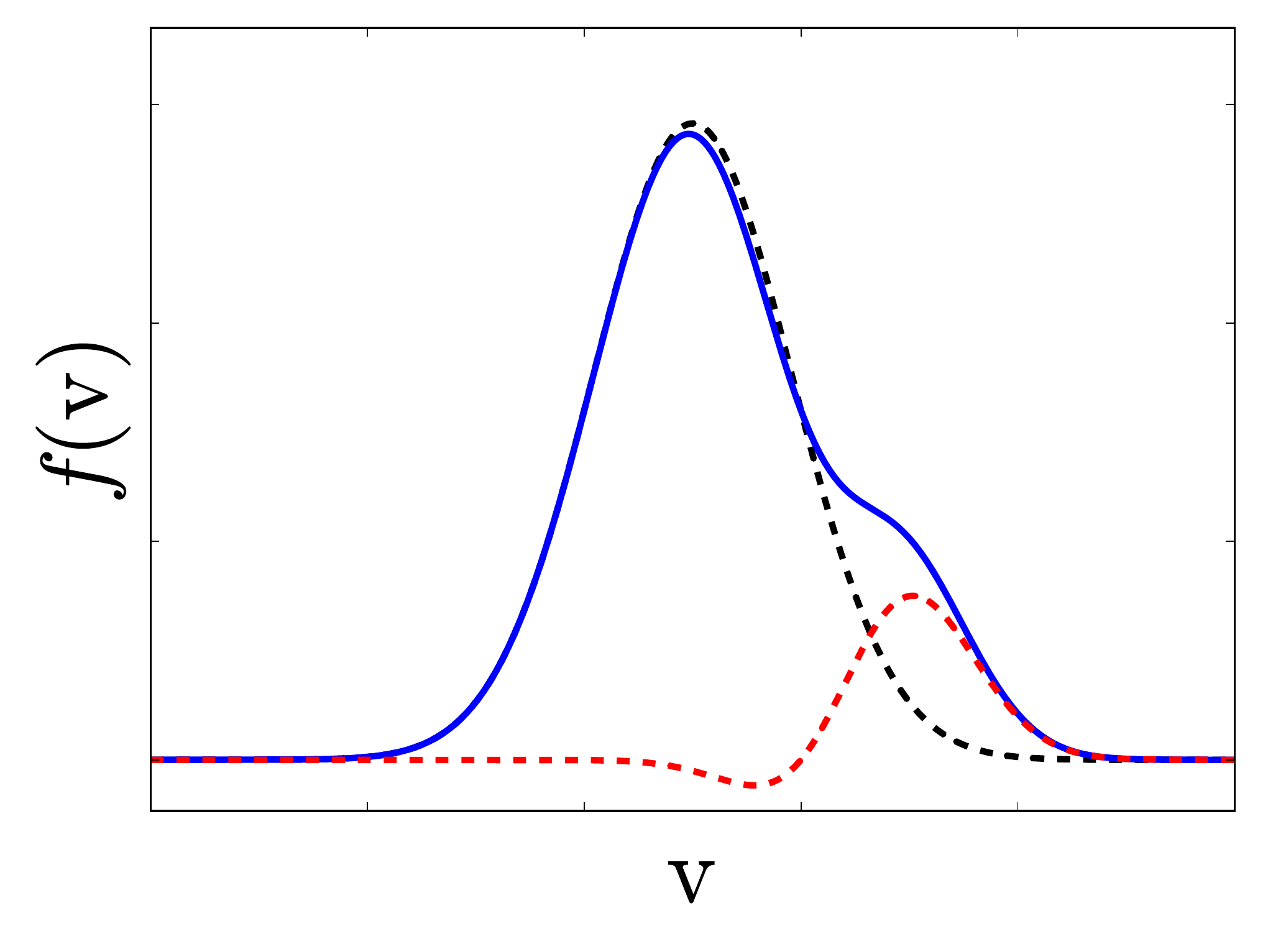}
\label{fig.fnM.Generalization.AsymTail2}
\end{subfigure}%
\caption{A possible asymmetric correction of INMDFs given by \refeq{eq.fnM.Generalization}.}
\label{fig.fnM.Generalization.AsymTail}
\end{figure}

It is possible to use other formulas for the description of non-Maxwellians such as shown in
\begin{figure}[!ht]
\centering
\begin{subfigure}[]{}
\includegraphics[width=40mm,natwidth=585,natheight=441]{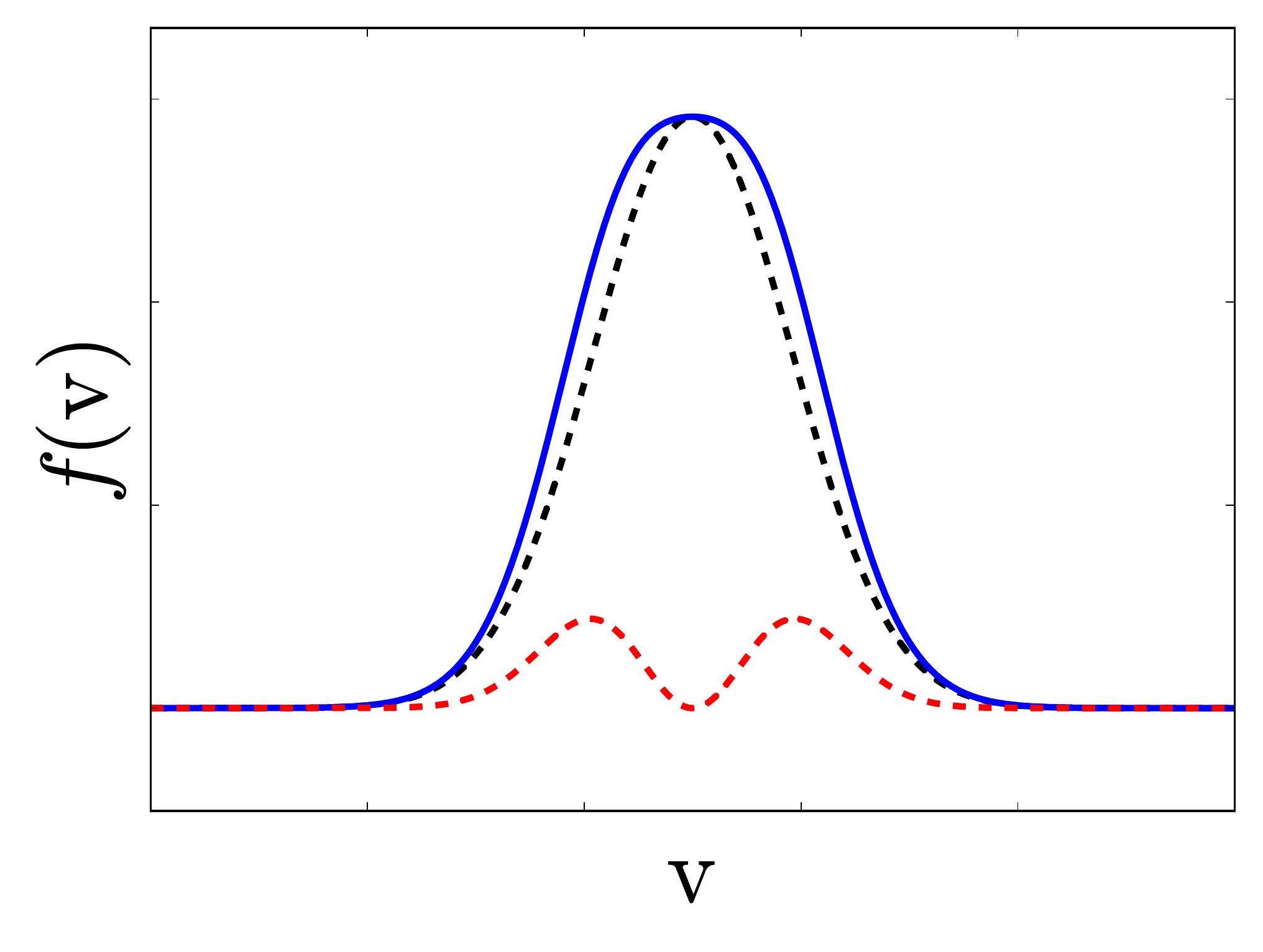}
\label{fig.Izacardian_examples2.1}
\end{subfigure}%
\begin{subfigure}[]{}
\includegraphics[height=30mm,natwidth=585,natheight=441]{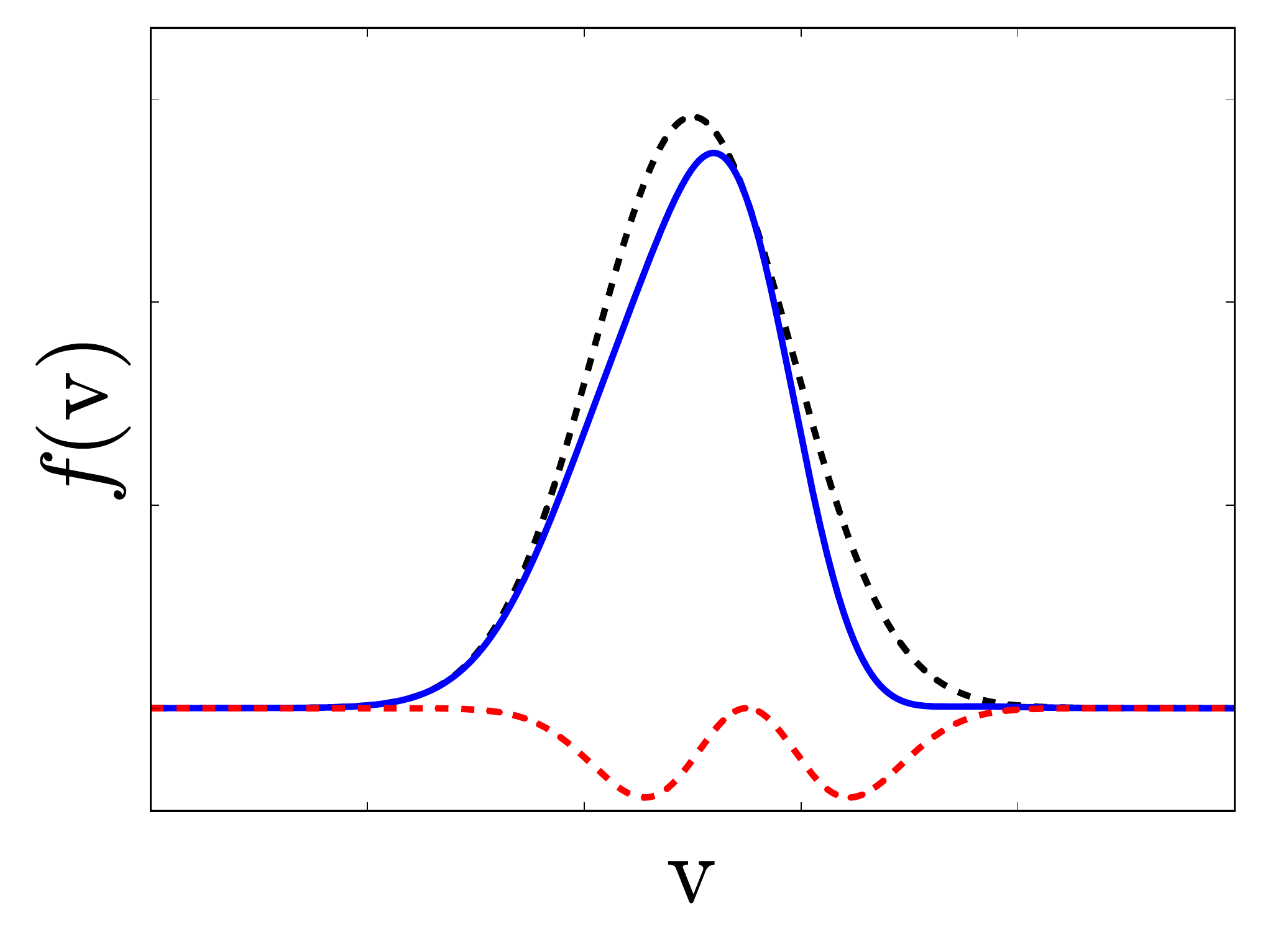}
\label{fig.Izacardian_examples2.4}
\end{subfigure}%
\caption{Example of two INMDFs obtained from \refeq{eq.fnM.Generalization} with different values of the fluid hidden variables.}
\label{fig.Izacardian_examples2}
\end{figure}
Fig.~\ref{fig.Izacardian_examples2} where different NMDFs are obtained from
\bgeqa
\label{eq.fnM.Generalization}
\displaystyle
f &=& \sum_{k=0}^{N_k} \frac{ a_k \left( {\rm v} - b_k \right)^{n_k} }{\left(2\pi d_k^{m_k}\right)^{1/2}}  \exp\left( -\frac { \left( {\rm v}-c_k\right)^2 }{ 2e_k }  \right),
\edeqa
with $m_k = 2 {\rm E}[(n_k+1)/2]+1$, $n_k \in \mathbb{N}$ where ${\rm E}[x]$ is the floor function and $a_k$, $b_k$, $c_k$, $d_k$, $e_k$ are fluid coefficients for all integer $k$. The set of coefficients $\{ a_k, b_k, c_k, d_k, e_k \}$ are the hidden variables. We notice that $a_k$ has a dimension of a density if $n_k$ is even and of a particle flux if $n_k$ is odd, $b_k$ and $c_k$ have the dimension of velocities and $d_k$ and $e_k$ of temperatures. This general form can describe asymmetric distribution functions $f$ which deviate from MDFs $f_0$ and the effective density can be different to $n$ (e.g., when we consider odd values of $n_k$). Moreover, because the hidden variables $b_k$ and $c_k$ as well as $d_k$ and $e_k$ can be different, in opposite to the Fig.~\ref{fig.Izacardian_examples1}, we can consider asymmetric corrections $\delta f$ as shown by red curves in Fig.~\ref{fig.fnM.Generalization.AsymTail}.
Further investigations using this generalized INMDFs given by \refeq{eq.fnM.Generalization} are possible. In summary, this formulation (i.e., the generalized INMDFs) can be related to a generalization of the Hermite polynomial representation~\cite{Grad:1949a,Grad:1949b,Grad:1963} (due to the presence of $({\rm v}-b_k)^{n_k}$ term) but each term is associated with a different MDF. The representation given by \refeq{eq.fnM.Generalization} forms a non-orthogonal basis in opposition to usual work found in the literature where orthogonal basis are used in order to project the distribution function on the basis set, but here this non-orthogonal basis is relevant to a significant reduction of the number of required hidden variables. Moreover, the completeness property is intrinsically inherited from the usual Hermite polynomial since \refeq{eq.fnM.Generalization} can be reduced to an Hermite polynomial with specific relation between all hidden variables (similar argument with a possible reduction to the RGBF, see below). The main advantage of this representation against the Hermite polynomial is that the required number of terms can be drastically reduced for the description of non-Maxwellians with high flows.  
Some researchers argue that one needs to keep at least hundreds or thousands of terms with the usual Hermite polynomial representation (so at least hundreds or thousands independent fluid moments, see Ref.~\cite{Izacard_2016_Paper2}) for an accurate description of non-Maxwellian plasmas. With our generalization given by \refeq{eq.fnM.Generalization}, it is conceivable to keep much less hidden variables (related to much less independent fluid moments). Having less hidden variables is an argument of the advantage of the INMDF. Another argument is the physical interpretation (see the end of Sec.~\ref{ToC.NMDFs.INMDF}) of the hidden variables. We remark that our general representation can also reproduce the sum of shifted MDFs (i.e., the RGBF) by imposing other constraints such as $n_k=0$ and $m_k=1$.\\
The moments $M_q$ of the generalized distribution function given by \refeq{eq.fnM.Generalization} become
\bgeqa
\label{eq.fnM.Generalization.Mq}
\displaystyle
M_q &=& \sum_{k=0}^{N_k} \Bigg[ \Delta_k \sum_{p=0}^{n_k} \left(\begin{array}{c}n_k\\p\end{array}\right) \left(-b_k\right)^{(n_k-p)} \Bigg. \nonumber\\
\Bigg. && \qquad\qquad\qquad
\times
J_{p+q}\left( \frac{1}{2e_k}, \frac{c_k}{e_k} \right) \Bigg],
\edeqa
where $\Delta_k = a_k \left(2\pi d_k^{m_k}\right)^{-1/2} \exp\left( - c_k^2 / ( 2e_k ) \right)$, $\left(\begin{array}{c}m\\n\end{array}\right)$ is the combination and the function $J_k(a,b)$ is introduced above. The introduced INMDF given by \refeq{eq.nM.fINMDF} or \refeq{eq.fnM.Generalization} is a new minimal formulation for a local non-Maxwellian deviation. This distribution function cannot be exactly obtained with a finite number of terms by any of the existing basis functions. In comparison to the sum of MDFs (i.e., the RGBF~\cite{Hirvijoki:2015}), valid only at two collisionality limits, the INMDF is the first distribution function which can be self-consistent with MDFs steady states at finite collisionality. More properties of the INMDF will be published later.

\subsection{Physical reality of the INMDFs}
\label{ToC.NMDFs.Reality}
The readers may be concerned about the physical reality of the newly introduced INMDFs with respect to the well observed KDF in astrophysics or the natural expansion with a sum of MDFs. In order to argue about the physical picture introduced by the displacement of population of particles in the velocity phase space of the INMDFs, we describe below a qualitative fitting of an indirect experimental measurement of the distribution function. In fact, to date there is no universal direct measurement of the velocity phase space variations of the distribution function but it is common to indirectly compute the distribution function from different diagnostics. Moreover, even if we choose to perform the qualitative fitting (without the necessity to use the experimental data) of only one published article on the electron temperature measurement discrepancy, it is at least possible (not shown here, see Ref.~\cite{Izacard_2016_Paper2}) to fit NMDFs observed by Fokker-Planck numerical codes in presence of lower hybrid current drive or by particle in cell codes in presence of ion orbit losses.
\begin{figure}[!htbp]
\centering
\includegraphics[height=50mm,natwidth=545,natheight=339]{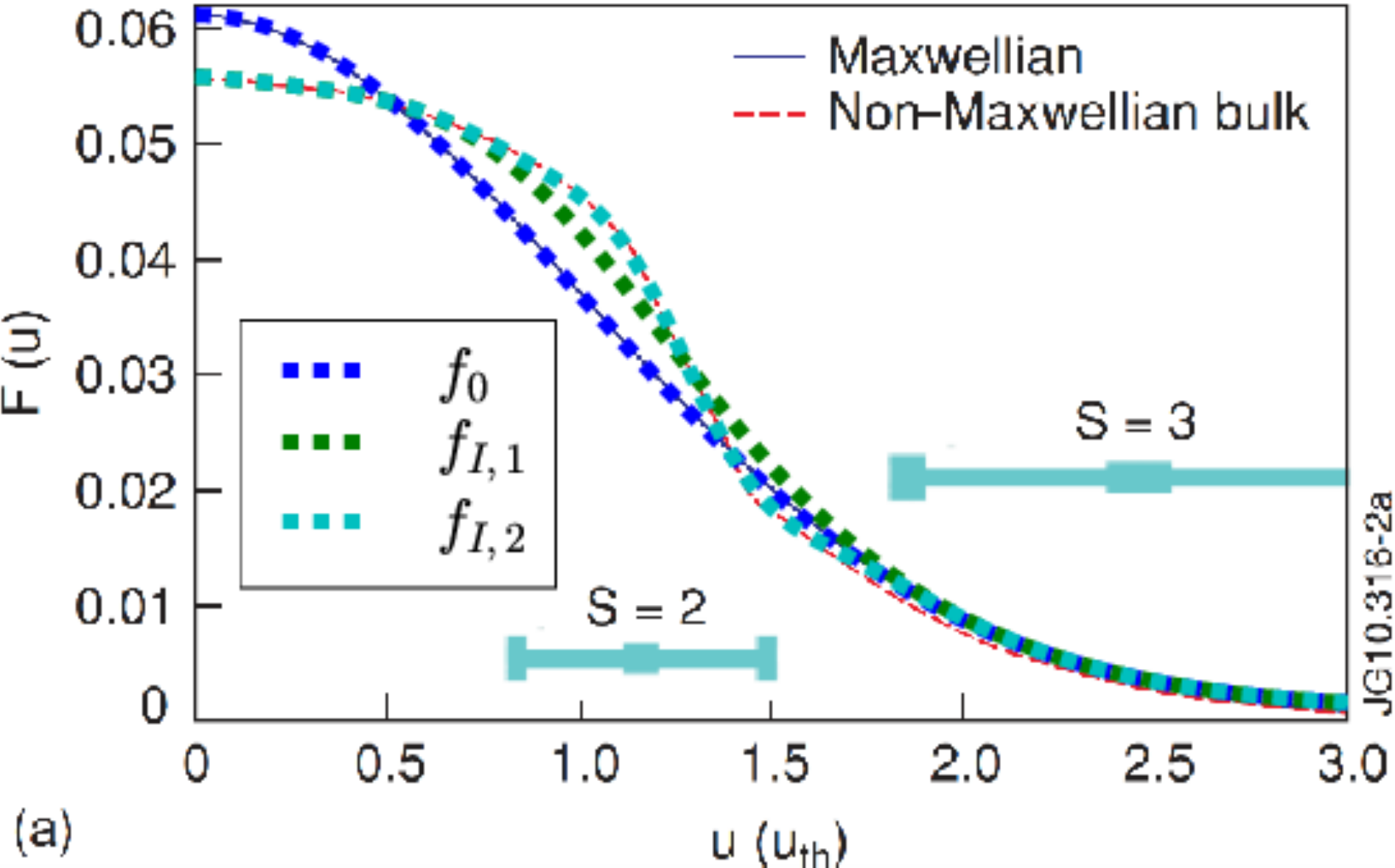}
\caption{Over plot of our qualitative fitting of a MDF $f_0$ (blue dashed cruve) and two INMDFs $f_{I,1}$ (green dashed curve) and $f_{I,2}$ (cyan dashed curve). Our fitting is done with respect to the numerical model NMDF (red dashed curve) of Ref.~\cite{Beausang_2011_RSI} which explains the inconsistency between the electron cyclotron emission and the Thomson scattering measurements. The background figure is reprinted from [Beausang, \textit{et al.}, \textit{Rev. Sci. Instrum.} \textbf{82}, 033514 (2011)] with the permission of AIP Publishing. See Fig.~2.a in Ref.~\cite{Beausang_2011_RSI}.}
\label{fig.Real.ECE-TS}
\end{figure}
Fig.~\ref{fig.Real.ECE-TS} proves the capability of INMDFs to describe experimentally observed physical processes. With some density and temperature constraints in presence of NBI and ICRF in JET~\cite{Luna_2003_RSI,Beausang_2011_RSI} and TFTR~\cite{Taylor_1996_PPCF,Fidone_1996_PoP,White_2012_NF}, it has been observed a $20\%$ discrepancy of the electron temperature between the interpretation from the electron cyclotron emission (ECE) and Thomson scattering (TS) due to kinetic effects~\cite{Beausang_2011_RSI}. De La Luna, \textit{et al.} and Beausang, \textit{et al.} numerically found a model NMDF (dashed red curve) constructed from the spectrum of the ECE. {In fact, the measurement of the ECE spectrum was not consistent with a MDF. De La Luna, \textit{et al.} originally found the way to better recover this ECE spectrum by numerically modifying the distribution function and obtaining the called model NMDF. It turned out that} this model NMDF resolves the discrepancy of the interpreted electron temperature by TS. The physical processes at the origin of this model NMDF are not understood. Our three over plots (blue, green and cyan dashed curves) in Fig.~\ref{fig.Real.ECE-TS} can help to understand the origin of the detected NMDF. The blue curve $f_0$ corresponds to a MDF with a specific set of parameters. The green curve $f_{I,1}=f_0+\delta f_1$ is obtained by adding $\delta f_1$ with the parameters $(q,r,s)=(1.7,0.52,0.48)$. Then, the cyan dashed curve $f_{I,2}=f_{I,1}+\delta f_2$ is obtained by adding $\delta f_2$ with $(q,r,s)=(-0.2,1.3,0.2)$. This means that we detect both, an enhancement (i.e., a heating $\delta f_1$) and a reduction (i.e., a cooling $\delta f_2$) of the energy of two different populations of particles. {We notice that because we use interpreted parameters, we do not need to access the data (i.e., the density, fluid velocity and temperature of the background MDF). Our additional parameters ($\Gamma,c,W$) are the only way to modify the shape of the distribution function and the use of the terms ($q,r,s$) make our figure independent of the data.} We prove here that the $20\%$ TS-ECE discrepancy is due to approximately $1.7\%$ of heated particles and $0.2\%$ of cooled particles, and both are attracted by a resonant-like process around $1.0\ u_{\rm th}$. {We highlight the fact that this result may not be the detection of a new physical effect but, at least, it helps to describe and understand the unsolved problem of the TS-ECE discrepancy. Future work could try to link our understandings with known heating and current drive resonant theories or may lead to the discovery of a new physical effect. Fokker-Planck codes may help to fully understand the details of the energy transfer. In summary, we do not attempt to describe here what is resonant with what in this complex unsolved problem where an electron NMDF, associated here with a kind of current drive resonance, is experimentally observed in presence of ion heating and current drive (NBI and ICRH). 
However, even if w}e do not know yet the exact origin of these detected physical processes (heating and cooling), the analytic formula of the INMDF $f_{I,2}$ can be directly used to understand the effects of non-Maxwellian bulks on other diagnostics and on the transport and turbulence. Indeed, even without theoretically or numerically describing the physical processes at the origin of the heating and cooling down we observed here, we can still have a good description of the deviations of the NMDF with respect to a MDF and analytically predict and validate the kinetic effects from this observed NMDF on different diagnostics. {Our capability to calibrate different experimental diagnostics is indispensable since we prove here that less than $2\%$ of super-thermal particles can lead to a $20\%$ discrepancy of the electron temperature interpretation between TS and ECE when a MDF is assumed. It is specially indispensable for the prediction of the effects of super-thermal tails on different diagnostics in ITER where a $90\%$ of self-heating by fusion reactions is planned in addition to some external heating.} Future work will focus on our detected INMDF $f_{I,2}$ and will help to initiate standard {experimental} procedures to measure NMDFs and to resolve inconsistencies between diagnostics more or less sensitives to kinetic effects. \\

In the next section, the advantage of using one of these NMDFs is shown for the analytic computation of kinetic corrections on the secondary electron emission, the Langmuir probe characteristic curve and the entropy.

\section{Kinetic corrections due to Non-Maxwellians}
\label{ToC.KinCorrections}
This section applies the NMDFs shown in the previous section on some existing theories relevant to current and future plasma devices. Analytic corrections of these theories are given as function of the fluid hidden variables. These corrections can be implemented in existing numerical simulations in order to better describe some kinetic effects when they need to be taken into account.

\subsection{Corrections for the secondary electron emission}
In tokamaks scrape-off-layer (SOL), incident primary particles (i.e., charged or neutral particles) release secondary electron from a solid surface. The secondary electron emission (SEE) reduces the sheath potential and can modify many phenomena in the plasma-material interface~\cite{Bruining:1954}. The computation of the SEE is obtained by formulating the SEE yield for incoming particles of energy $E$ related to the material under investigation (Sternglass formula~\cite{Bruining:1954,
Sternglass_1957_PR,
Dionne:1975}) multiplied by the proportion of particles at this energy (i.e., the distribution function $f(E)$) and has been under investigation in the literature~\cite{Meyer-Vernet_1982_AA,
Mayer_1988_PRB,
Bacharis_2010_PoP_17,
Raitses:2011,
Shotorban_2015_PRE}. By integrating over the possible energy of incoming particles, the secondary electron emission $\delta_{see}$ with respect to the incoming particles distribution function $f(E)$ becomes
\bgeqa
\label{eq.SEE}
\displaystyle
\delta_{see} = \frac{\int_0^{\infty} \sqrt{\frac{2E}{m}} f(E) \delta_{s}(E) dE}{\int_0^{\infty} \sqrt{\frac{2E}{m}} f(E) dE},
\edeqa
with $\delta_{s}(E) = \Delta_{s}\ E \exp\left( -2 \sqrt{E} / \sqrt{E_{\rm max}} \right)$, the constant coefficient $\Delta_{s} = (2.72)^2 \delta_{\rm max} / E_{\rm max}$, and $E=m{\rm v}^2/2$ is the kinetic energy. Some values of $E_{\rm max}$ and $\delta_{\rm max}$ have been obtained experimentally and are dependent of the material under consideration (e.g., see Refs.~\cite{Bacharis_2010_PoP_17,Dionne:1975}). The value of $E_{\rm max}$ corresponds to the energy associated to the value of $\delta_{\rm max}$, the maximum SEE measured and used to obtain a normalized $\delta_{see}$. The empirical formula for a MDF obtained in Ref.~\cite{Bacharis_2010_PoP_17} is
\bgeqa
\displaystyle
\log_{10}\left[ \delta_{see}(T) \right] \approx C_3 x^3 + C_2 x^2 + C_1 x + C_0,
\edeqa
with $x = \log_{10}(T)$, the coefficient $C_k$ are given ad-hoc for each species, and the dependence with respect to the fluid velocity $v$ disappears by using the assumption of a Maxwellian plasma with no mean flow. However, we can reformulate \refeq{eq.SEE} to
\bgeqa
\label{eq.SEE2}
\displaystyle
\delta_{see} = \frac{\int_0^{\infty} {\rm v}^2 f({\rm v}) \delta_{s}({\rm v}) d{\rm v}}{\int_0^{\infty} {\rm v}^2 f({\rm v}) d{\rm v}},
\edeqa
with $\delta_{s}({\rm v}) = \Delta_{s}\ m {\rm v}^2/2  \exp\left( -2{\rm v}/ u \right)$, and $u=\sqrt{2E_{\rm max}/m}$. \begin{figure}[!ht]
\centering
\begin{subfigure}[]{}
\centering
\includegraphics[height=65mm,natwidth=576,natheight=432]{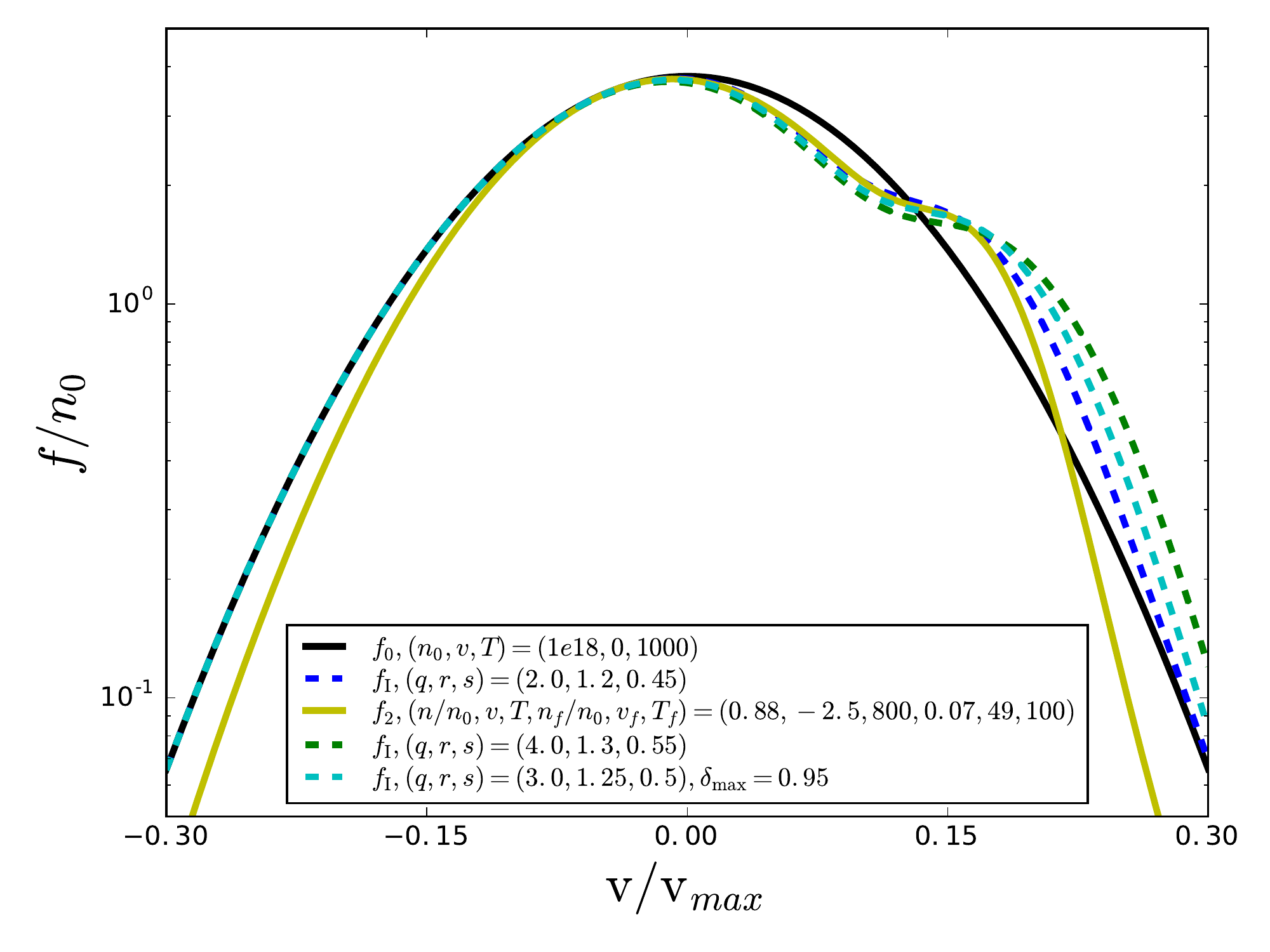}
\label{fig.fnM.SEE_DF}
\end{subfigure}%
\begin{subfigure}[]{}
\centering
\includegraphics[height=65mm,natwidth=576,natheight=432]{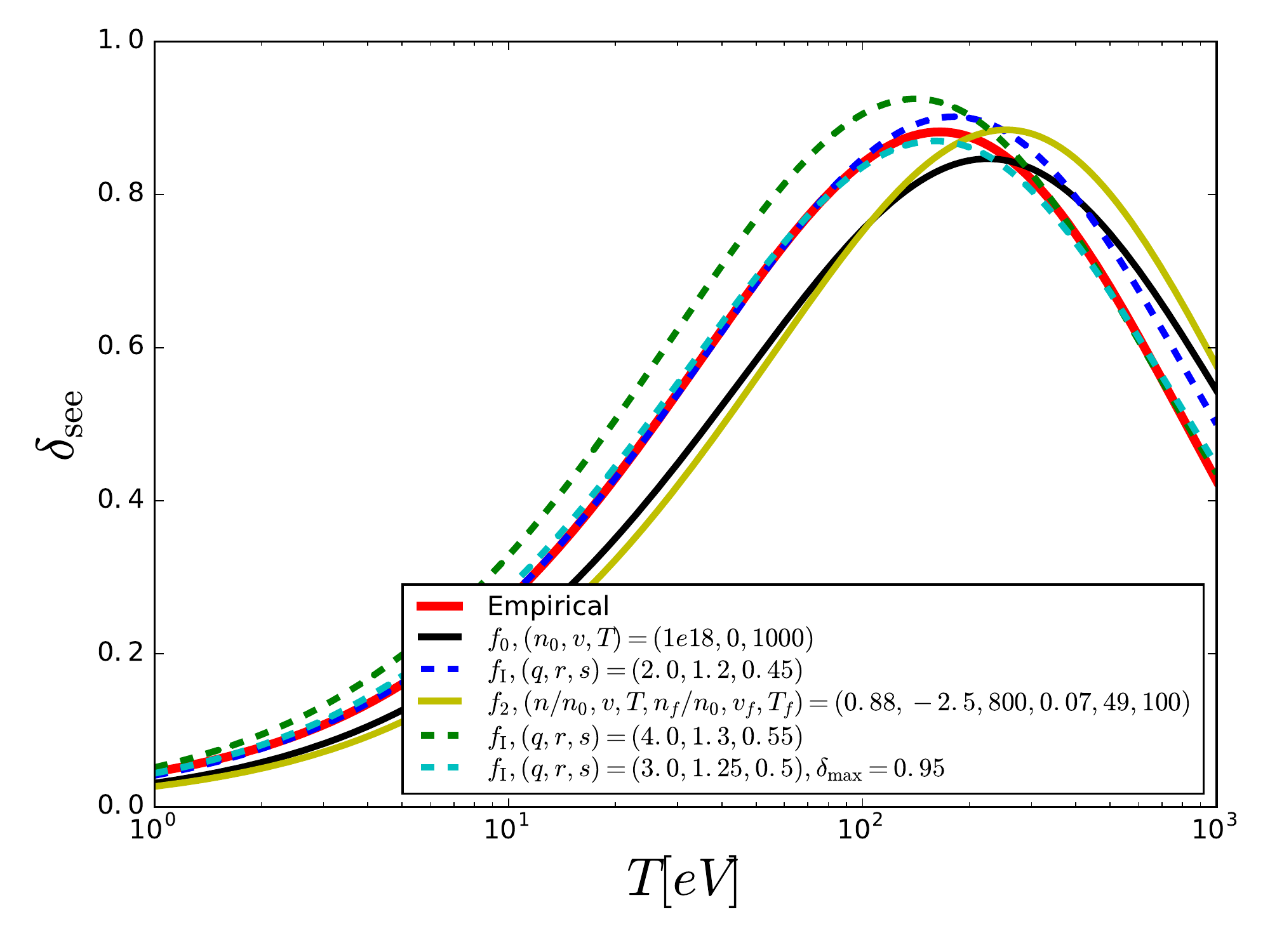}
\label{fig.fnM.SEE_Modif}
\end{subfigure}%
\caption{Distribution functions (a) and effects of the tails on the SEE (b) from \refeq{eq.SEE.dsec0} for a Maxwellian and from Eqs.~(\ref{eq.SEE.dsec2M}) and (\ref{eq.SEE.dsecI}) for different INMDFs.}
\label{fig.fnM.SEE}
\end{figure}
Here we found the following exact analytic formula from \refeq{eq.SEE2} for a MDF given by \refeq{eq.nM.fINMDF.f0}
\bgeqa
\label{eq.SEE.dsec0}
\displaystyle
\delta_{see,0}(v,T) = \delta \frac{J_{4}\left(\frac{1}{2T},\frac{v}{T}-\frac{2}{u},0\right)}{J_{2}\left(\frac{1}{2T},\frac{v}{T},0\right)},
\edeqa
with $\delta = (2.72)^2 / (8 \sqrt{\pi}) \delta_{\rm max} / u^2$ and where the general form and some functions $J_k(a,b,0)$ are detailed in App.~\ref{ToC.App.Jkab0} where
\bgeqa
\label{eq.SEE.Jkab0}
\displaystyle
J_k(a,b,0) = \int_0^{\infty} {\rm v}^k \exp\left(-a{\rm v}^2+b{\rm v}\right) d{\rm v}.
\edeqa
{Following the same analytic method, it has been possible to obtain the corrections $\delta_{see,\kappa}$ from a KDF. However, we do not focus on this result here because }
it involves the hypergeometric PFQ function instead of $K_q(\kappa,T)$ due to of the presence of $\exp(-2 {\rm v} / u)$ in the term $\delta_{s}({\rm v})$ and it may not be possible to obtain a simple analytic form in comparison to the following results. This may mean that the KDF is less natural than the sum of two MDFs or the INMDF (see results below) due to the complexity of the solution of the SEE $\delta_{see,\kappa}$. We found for the sum of two Maxwellians the following corrections
\bgeqa
\label{eq.SEE.dsec2M}
\displaystyle
\delta_{see,2M} &=& \delta \Bigg[ \Delta J_{4}\left(\frac{1}{2T},\frac{v}{T}-\frac{2}{u},0\right) \Bigg. \nonumber\\
\displaystyle
\Bigg. &&\quad
+ \Delta_f J_{4}\left(\frac{1}{2T_f},\frac{v_f}{T_f}-\frac{2}{u},0\right)  \Bigg] \nonumber\\
\displaystyle
&&
\times \Bigg[ \Delta J_{2}\left(\frac{1}{2T},\frac{v}{T},0\right) \Bigg. \nonumber\\
\displaystyle
\Bigg. &&\quad
+ \Delta_f J_{2}\left(\frac{1}{2T_f},\frac{v_f}{T_f},0\right) \Bigg]^{-1},
\edeqa
and for the INMDF given by \refeq{eq.nM.fINMDF} the following corrections
\bgeqa
\label{eq.SEE.dsecI}
\displaystyle
\delta_{see,I} &=& \delta \Bigg[ \Delta J_{4}\left(\frac{1}{2T},\frac{v}{T}-\frac{2}{u},0\right) \Bigg. \nonumber\\
\displaystyle
\Bigg. &&\quad
+ \Delta_I J_{5}\left(\frac{1}{2W},\frac{c}{W}-\frac{2}{u},0\right) \Bigg. \nonumber\\
\displaystyle
\Bigg. &&\quad
- c \Delta_I J_{4}\left(\frac{1}{2W},\frac{c}{W}-\frac{2}{u},0\right)   \Bigg] \nonumber\\
\displaystyle
&&
\times \Bigg[ \Delta J_{2}\left(\frac{1}{2T},\frac{v}{T},0\right) + \Delta_I J_{3}\left(\frac{1}{2W},\frac{c}{W},0\right) \Bigg. \nonumber\\
\displaystyle
\Bigg. &&\quad
- c \Delta_I J_{2}\left(\frac{1}{2W},\frac{c}{W},0\right)   \Bigg]^{-1},
\edeqa
where the secondary electron emissions $\delta_{see,0}(v,T)$, $\delta_{see,2M}(n,v,T,n_f,v_f,T_f)$ and $\delta_{see,I}(n,v,T,\Gamma,c,W)$ are functions of all fluid hidden variables and the quantities $\delta$, $\Delta$, $\Delta_f$ and $\Delta_I$ are previously defined.\\
Fig.~\ref{fig.fnM.SEE}.(a) shows different NMDFs computed from the bi-modal NMDF given by \refeq{eq.nM.f2Maxw} and from the INMDF given by \refeq{eq.nM.fINMDF} for a temperature $T=1keV$. The corrections of the secondary electron emission as function of the temperature are shown in Fig.~\ref{fig.fnM.SEE}.(b). \\ 
An important result is that the black curve obtained for a MDF with our exact formula does not match exactly the red curve of the empirical formula given in Ref.~\cite{Bacharis_2010_PoP_17}. The reason is due to an undetected presence of super-thermal particles in Ref.~\cite{Bacharis_2010_PoP_17} because our analytic formula $\delta_{see,I}$ with the INMDF $f_I(q=3,r=1.25,s=0.5)$ (cyan dashed curve) matches better the empirical formula (red curve) better in opposition to our analytic formula $\delta_{see,0}$ with the MDF $f_0(n_0=10^{18},v=0,T=10^3)$ (black curve). Moreover, the empirical formula is better matched for $T<10^2eV$ with $f_I(q=2,r=1.2,s=0.45)$ (blue dashed curve) and for $T>3.10^2eV$ with $f_I(q=4,r=1.3,s=0.55)$ (green dashed curve). We remark that only for the cyan curve in Fig.~\ref{fig.fnM.SEE}.(b) we used $\delta_{\rm max}=0.95$ instead of $1$, but it is possible to find other parameters which match the red empirical curve (e.g., keeping $\delta_{\rm max}=1$ and multiplying $n$ and $\Gamma$ by 0.95 because $\delta_{see}$ is linear with respect to these two quantities). Finally, it seems that the super-thermal particles of the empirical formula are clearly better described by the INMDF rather than by a sum of two MDFs (yellow solid curve is one of the best fit). Nevertheless, the goal is to predict the modifications of the secondary electron emission in presence of different super-thermal particles. The effects of a tail on the secondary electron emission are significant specially for a low temperature (i.e., $T \leq 10eV$) of the background plasma since a factor around $5$ can be observed (i.e., $(q,r,s)=(2,2.5,1)$ not shown here). Then, it is indispensable to take into account super-thermal particles even if this population represents less than $4\%$ because, at least, the dynamics of dusts, neutrals and impurities are highly impacted by the secondary electron emission. It looks like our analytic prediction (with $s>1$) is able to observe two peaks in the SEE like it has been experimentally observed (see Ref.~\cite{Raitses:2015}). Future work will investigate this two peaks observation, particularly the addition of other terms in \refeq{eq.nM.fINMDF}, in order to avoid negative distribution function for all ${\rm v}<c$.

\subsection{Corrections for the Langmuir probes interpretations}
A commonly used diagnostic of edge (cold) plasma is the Langmuir probe. By applying an electric potential scan on the Langmuir probe, we can measure the current of the probe as function of the applied potential. This give us the well known characteristic curve. The theory of the Langmuir probe has successfully been developed in the past century~\cite{Stangeby_2000,Hutchinson_2002} and properties of the plasma can be extracted from the characteristic curves such as the electron temperature, or the floating and plasma potentials by interpreting the measurement for a background Maxwellian plasma. However, a recurrent discrepancy of the interpretation of the plasma properties with other diagnostics (e.g., with Thomson scattering measurements~\cite{Luna_2003_RSI,
Beausang_2011_RSI,
Jaworski_2015_Private} in attached plasma) has been observed. This discrepancy has been related to NMDFs as detailed in Refs.~\cite{Popov:2009,Jaworski_2012_FED,Jaworski:2013} where bi-modal distribution functions (distribution function of $2$ populations of particles at different temperatures) have been used. Other results are published in the literature describing effects of non-Maxwellians on the Langmuir probe measurements~\cite{Arslanbekov,Sudit_1994_JAP,
Stangeby_1995_PPCF,
Batishchev_1997_PoP,
Demidov_2002_RSI,
Tskhahaya_2011_JNM,
Sheehan_2011_PoP,
Godyak_2011_JPD,
Jaworski_2012_FED,
Popov_2012_PSST,
Popov_2014_CPP,
Godyak_2015_JAP}. The bi-modal approximation is the first efficient way to describe NMDFs, but it assumes the superposition of two populations of particles at the Maxwellian equilibrium and the self-consistency is very restricted since in this bi-modal description the plasma is enough collisional to assure a MDF for each species but do not allow collisions between these two populations. In order to resolve this inconsistency, we propose to use the INMDF of the plasma (only one population of particles with $1$ effective temperature) which contains super-thermal particles. In our case, all particles (viewed as one species) collide together and external sources or sinks of energy create non-Maxwellian steady states. \\
The general formula of the electron current~\cite{Arslanbekov,Demidov_2002_RSI,Jaworski_2012_FED,Popov_2014_CPP} can be written as a function of any distribution function $f(E)$ of the kinetic energy $E$ of particles
\bgeqa
\label{eq.LP.Ie.E}
\displaystyle
I_e(U) = - \frac{8\pi e S}{3 m^2} \int_{eU}^{\infty} \frac{(E-eU) f(E)}{\gamma(E) \left[ 1 + \frac{E-eU}{E} \psi(E) \right]} dE,
\edeqa
where $\psi(E)$ is a diffusion parameter, $\gamma(E)$ a geometric parameter of the probe, $U=U_p-U_{pl}$ is the difference between the applied potential at the probe and the plasma potential, $S$ the surface of the probe, and $m$ is the mass of electron. The classical regime is obtained by assuming a diffusionless limit ($\psi(E) \ll 1$), and with $\gamma=4/3$ for a spherical probes. This formula computes the flux of electron which have higher relative energy than the potential $eU$ because electrons of lower energy cannot contribute to the current due to the Coulomb barrier generated by other electrons at the probe interface. With the change of coordinate $E=m{\rm v}^2/2$ and $u=\sqrt{2eU/m}$, the electron current becomes
\bgeqa
\label{eq.LP.Ie.v}
\displaystyle
I_e(U) = - \frac{2\pi e S}{m} \int_{u}^{\infty} \left(\frac{m{\rm v}^2}{2}-eU\right) {\rm v} f({\rm v}) d{\rm v}.
\edeqa
If we assume a background MDF $f_0$ given by \refeq{eq.nM.fINMDF.f0}, the electron current becomes
\bgeqa
\label{eq.LP.Ie0}
\displaystyle
I_{e,0}(U) &=& - \Delta \frac{2\pi e S}{m} \bigg( \frac{m}{2} J_3\left(\frac{1}{2T},\frac{v}{T},u\right) \bigg. \nonumber\\
\displaystyle
\bigg. && \qquad\qquad\quad
- eU J_1\left(\frac{1}{2T},\frac{v}{T},u\right) \bigg),
\edeqa
with the definition of the function
\bgeqa
\label{eq.LP.Jkabc}
\displaystyle
J_k(a,b,c) = \int_{c}^{\infty} {\rm v}^k \exp\left(-a{\rm v}^2+b{\rm v}\right) d{\rm v},
\edeqa
detailed in App.~\ref{ToC.App.Jkabc}. 
Then, from the KDF given by \refeq{eq.nM.fKappa} the electron current becomes
\bgeqa
\label{eq.LP.IeKappa}
\displaystyle
I_{e,\kappa}(U) &=& - \Delta_{\kappa} \frac{2\pi e S}{m} \bigg( \frac{m}{2} K_3\left(\kappa,T,u\right) \bigg. \nonumber\\
\displaystyle
\bigg. && \qquad\qquad\quad
- eU K_1\left(\kappa,T,u\right) \bigg),
\edeqa
from the sum of two MDFs given by \refeq{eq.nM.f2Maxw} the electron current becomes
\bgeqa
\label{eq.LP.Ie2M}
\displaystyle
I_{e,2M}(U) &=& - \frac{2\pi e S}{m} \bigg( \frac{m}{2} \Delta J_3\left(\frac{1}{2T},\frac{v}{T},u\right) \bigg. \nonumber\\
\displaystyle
&& \qquad\quad
- eU \Delta J_1\left(\frac{1}{2T},\frac{v}{T},u\right) \nonumber\\
\displaystyle
&&\qquad\quad
+ \frac{m}{2} \Delta_f J_3\left(\frac{1}{2T_f},\frac{v_f}{T_f},u\right) \nonumber\\
\displaystyle
&&\qquad\quad
- eU \Delta_f J_1\left(\frac{1}{2T_f},\frac{v_f}{T_f},u\right) \bigg),
\edeqa
and from the INMDF $f_I$ given by \refeq{eq.nM.fINMDF} the electron current becomes
\bgeqa
\label{eq.LP.IeI}
\displaystyle
I_{e,I}(U) &=& - \frac{2\pi e S}{m} \bigg( \frac{m}{2} \Delta J_3\left(\frac{1}{2T},\frac{v}{T},u\right) \bigg. \nonumber\\
\displaystyle
&& \qquad\quad
- eU \Delta J_1\left(\frac{1}{2T},\frac{v}{T},u\right) \nonumber\\
\displaystyle
&&\qquad\quad
+ \frac{m}{2} \Delta_I J_4\left(\frac{1}{2W},\frac{c}{W},u\right) \nonumber\\
\displaystyle
&&\qquad\quad
- \frac{m}{2} c \Delta_I J_3\left(\frac{1}{2W},\frac{c}{W},u\right) \nonumber\\
\displaystyle
&&\qquad\quad
- eU \Delta_I J_2\left(\frac{1}{2W},\frac{c}{W},u\right) \nonumber\\
\displaystyle
&&\qquad\quad
+ eUc \Delta_I J_1\left(\frac{1}{2W},\frac{c}{W},u\right) \bigg).
\edeqa
\begin{figure}[!ht]
\centering
\begin{subfigure}[]{}
\centering
\includegraphics[height=65mm,natwidth=576,natheight=432]{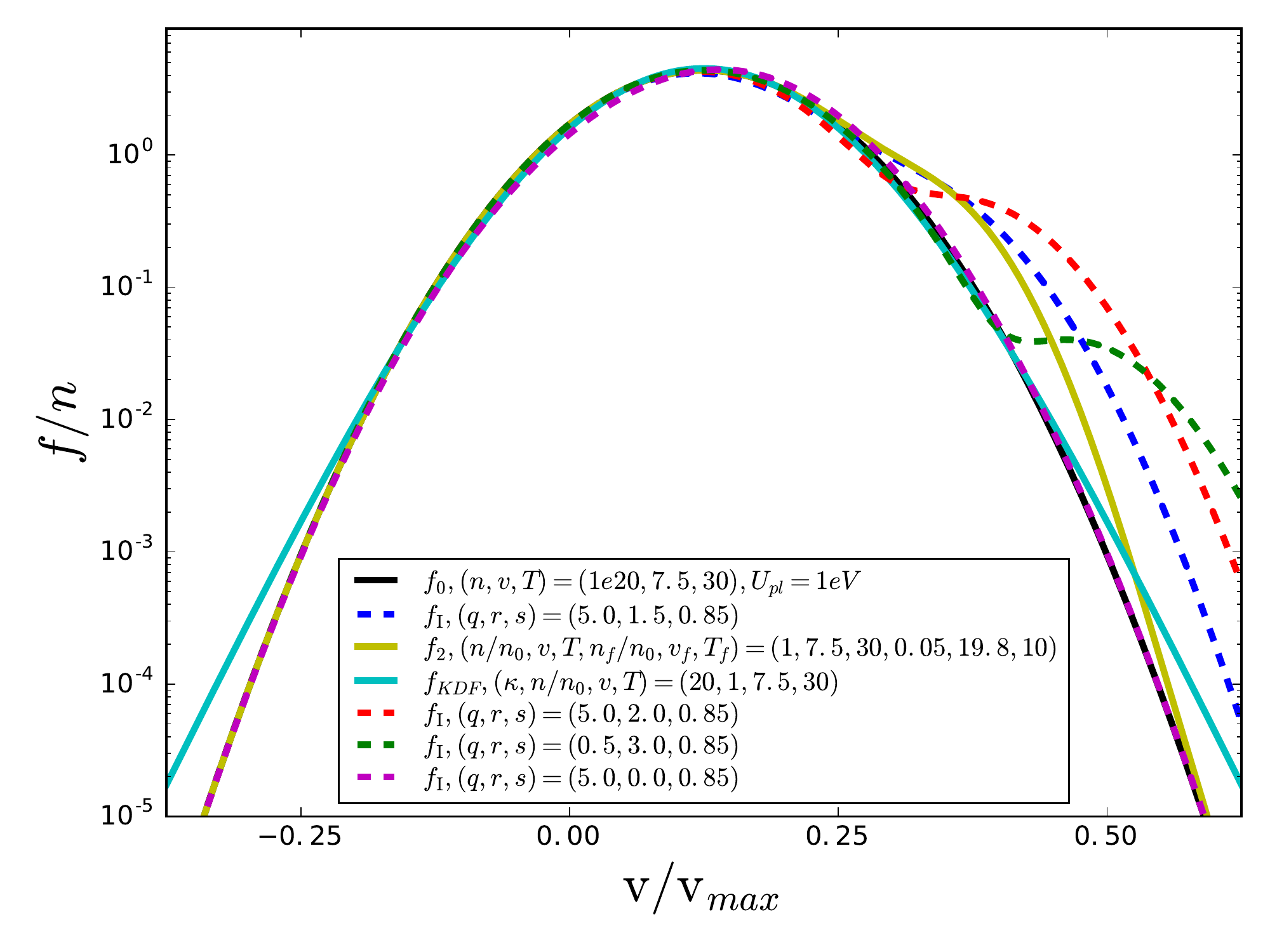}
\label{fig.fnM.LangmuirProves_DF}
\end{subfigure}%
\begin{subfigure}[]{}
\centering
\includegraphics[height=65mm,natwidth=576,natheight=432]{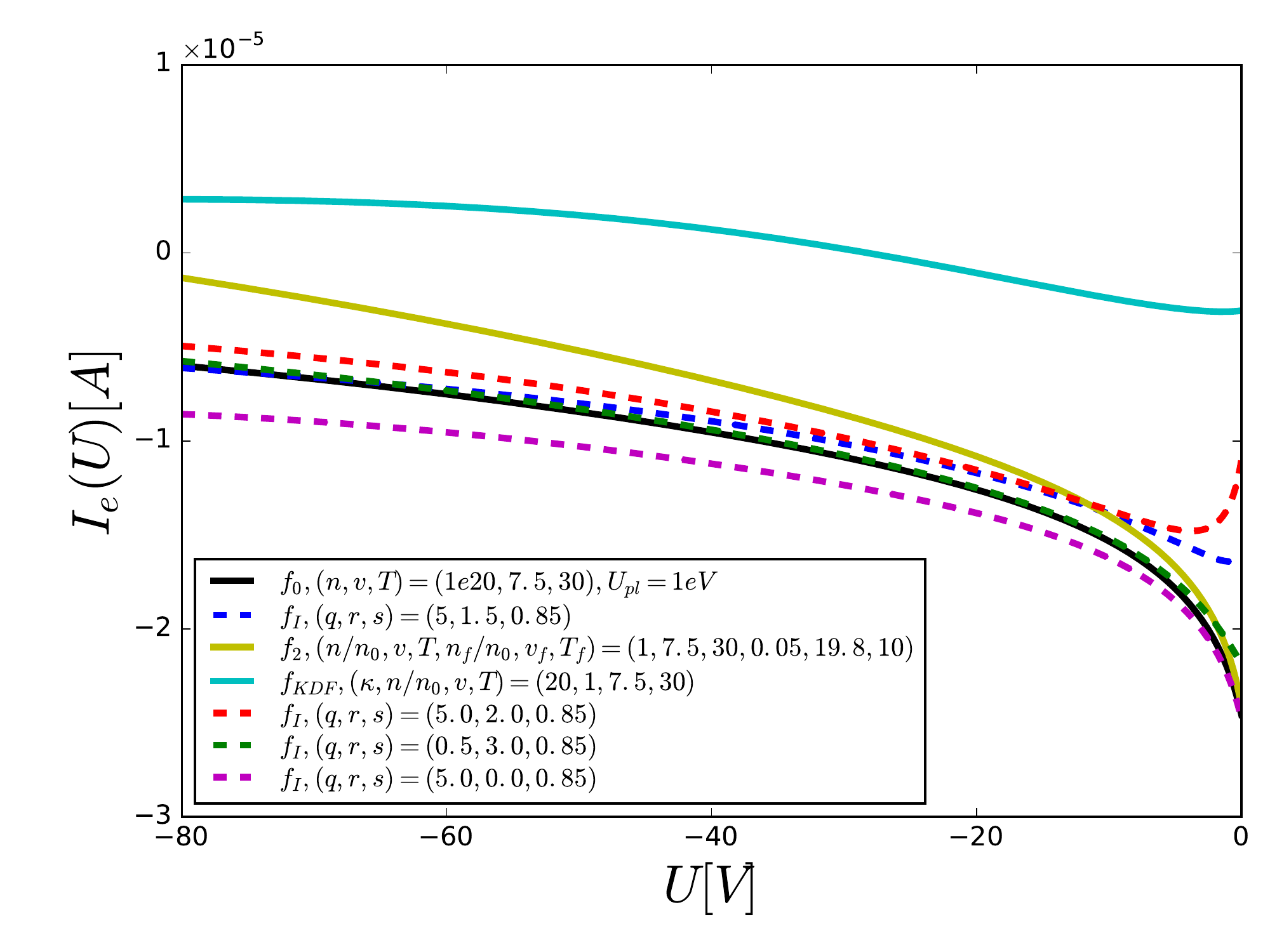}
\label{fig.fnM.LangmuirProbe_CharacteristicCurve}
\end{subfigure}%
\caption{Logarithms of distribution functions (a) and effects of the tails on the Langmuir probe current (b) from \refeq{eq.LP.Ie0} for a Maxwellian and \refeq{eq.LP.IeI} for different INMDFs.}
\label{fig.fnM.LangmuirProbe}
\end{figure}
As shown in Fig.~\ref{fig.fnM.LangmuirProbe}, different NMDFs (see Fig.~\ref{fig.fnM.LangmuirProbe}.(a)) with the same Maxwellian background (black curve) can significantly modify (see Fig.~\ref{fig.fnM.LangmuirProbe}.(b)) the electric current measured by a Langmuir probe. This figure is obtained for parameters relevant to the SOL plasmas close to the divertor plates (i.e., $T=15eV$, $n=10^{20} m^{-3}$). However, we remark that the amplitudes are not consistent with the experimental measurements by the Langmuir probes since the formula given by \refeq{eq.LP.Ie.E} (see Refs.~\cite{Arslanbekov,Demidov_2002_RSI,Jaworski_2012_FED,Popov_2014_CPP}) is not dimensionally correct. Usually, results are given in arbitrary units in the literature~\cite{Jaworski_2015_Private} in order to avoid this dimensional inconsistency. This dimensional issue can be investigated later since it is not the focus here. \\
Moreover, \refeq{eq.LP.Ie.v} (directly linked to the Druyvesteyn formula~\cite{Arslanbekov,Demidov_2002_RSI,Popov_2012_PSST,Popov_2014_CPP}) is the diffusionless limit of \refeq{eq.LP.Ie.E}. In comparison to other studies~\cite{Jaworski_2015_Private}, the INMDF used in the diffusionless limit \refeq{eq.LP.Ie.v} can reproduce similar curves than the use of a MDF in the diffusional \refeq{eq.LP.Ie.E}. This observation is not surprising because the particles diffusion describes an enhancement of the displacement of particles as well as the population of super-thermal particles. This is a very important observation because instead of using ad-hoc diffusion parameter in order to reproduce experimental observations, the existence of NMDFs could replace those. 

Finally, all common interpretations of the plasma parameters are modified by the NMDF. For example, the floating potential $U_f$ is defined by $I_e(U_f)=I_i(U_f)$ where $I_i(U)$ is the current of the Langmuir probe generated by ions and commonly assumed to be constant due to ion mass and temperature involved. From the result given by \refeq{eq.LP.IeI} and its version for the ion current, it is possible to better evaluate the floating potential from the distribution function. Of course, all other quantities such as the plasma potential $U_{pl}$, the electron temperature $T_e$ and the radial electric field $E_r$ are interpreted differently due to NMDFs. Additional investigations will be reported later.

\subsection{Entropy decrease}
In the thermodynamics theory, the entropy is interpreted as the degree of disorder of a system. A generalization of the entropy is required for the description of non thermodynamics equilibrium. By sharing similar perspectives than Refs.~\cite{Kranys_1970_ARMA,Livadiotis_2014_Entropy}, corrections of the entropy can be analytically computed from a NMDF. In an isolated plasma, the entropy can only increase due to collisions in order to reach its maximum value when the distribution function is Maxwellian. The definition of the statistical entropy developed by Boltzmann is
\bgeqa
\label{eq.B.s}
\displaystyle
s &=& - k_B \int_{-\infty}^{\infty} \frac{f}{n} \log\left( \frac{f}{n} \right) d{\rm v},
\edeqa
where $f$ is the distribution function of the particles of density $n=\int f d{\rm v}$ and $k_B$ is the Boltzmann constant. The maximal value of the entropy is obtained with a MDF $f_0$ (see App.~\ref{ToC.App.Entropy})
\bgeqa
\displaystyle
s_0 &=& - k_B \int_{-\infty}^{\infty} \frac{f_0}{n} \log\left( \frac{f_0}{n} \right) d{\rm v}, \\
\displaystyle
&=& \frac{k_B}{2} \big[ 1 + \log(2\pi T) \big],
\edeqa
such that $\partial_t s_0 = - k_B \partial_t \int f_0/n \log\left( f_0/n \right) d{\rm v} = k_B/(2T) \partial_t T= 0$ for isolated equilibrium plasmas. For other distribution functions than the MDF, the entropy can only be smaller to this value and increase ($\partial_t s \geq 0$). However, in order to make a link with observed NMDFs for non-isolated plasmas, it makes sense to interpret the entropy by a level of sharing energy or information. The maximal entropy obtained with a MDF of an isolated plasma is reached when the interaction between all particles of the plasmas have shared the total energy of the system after a thermalization characteristic time $t_{\rm th}$. This level of sharing can decrease when an external source or sink of energy of a non-isolated plasma appears locally in the velocity phase space at a finite collisionality. The presence of a steady-state super-thermal population of particles due to external sources introduces corrections of the entropy because at finite collisionality some particles can be sensitive to the external source of energy in time range $t < t_{\rm th}$, shorter than the one associated to the sharing of energy. The analytic computation of the corrections becomes possible using the INMDF written in the form $f_I = f_0 +\delta f = f_0 ( 1 + \delta f/f_0)$ with the MDF $f_0$ and the corrections $\delta f$ previously defined (respectively by Eqs.~(\ref{eq.nM.fINMDF.f0}) and~(\ref{eq.nM.fINMDF.delf})). The modified entropy due to the presence of a super-thermal population reads
\bgeqa
\displaystyle
s_I &=& - k_B \int_{-\infty}^{\infty} \frac{f_I}{n_I} \log\left( \frac{f_I}{n_I} \right) d{\rm v},
\edeqa
where $n_I=\int f_I d{\rm v} = n$. Using the Taylor expansion of the logarithm of $(1+x)$ around $0$ (i.e., $\log(1+x)=- \sum_{k=1}^{\infty} (-1)^k x^k / k$) if we assume relatively small deviations from the Maxwellian (i.e., $\vert \delta f \vert \ll f_0$), then, following the details of App.~\ref{ToC.App.Entropy}, the entropy becomes $s_I = s_0 + \delta s$ with
\bgeqa
\displaystyle
&&\delta s = - k_B \frac{1}{n} \int_{-\infty}^{\infty} \delta f \log\left( \frac{f_0}{n} \right) d{\rm v} \nonumber\\
\displaystyle
&&+ k_B \frac{\Delta}{n} \sum_{k=1}^{\infty} \frac{(-1)^k}{k(k+1)} \left( \frac{\Delta_I}{\Delta} \right)^{k+1} \sum_{m=0}^{k+1} (-c)^{k+1-m} \nonumber\\
\displaystyle
&&\times \left(\begin{array}{c}k+1\\ m\end{array}\right) J_m\left( \frac{k+1}{2W}-\frac{k}{2T}, \frac{(k+1)c}{W}-\frac{kv}{T} \right). \qquad
\edeqa
At the first order in the expansion of $\vert \delta f \vert \ll f_0$ (i.e., the sum is of higher order, see App.~\ref{ToC.App.Entropy}) and with the use of the following relations
\bgeqa
\displaystyle
J_1\left(A,B\right) - c J_0\left(A,B\right) &=& 0, \\
\displaystyle
J_2\left(A,B\right) - c J_1\left(A,B\right) &=& W J_0\left(A,B\right), \\
\displaystyle
J_3\left(A,B\right) - c J_2\left(A,B\right) &=& c (W-c^2) J_0\left(A,B\right), \\
\displaystyle
\Delta_I J_0\left(A,B\right) &=& \frac{\Gamma}{W},
\edeqa
when $A=1/(2W)$, $B=c/W$ and with the functions $J_k(a,b)$ given in App.~\ref{ToC.App.Jkab}, the correction of the entropy is
\bgeqa
\displaystyle
\delta s &\approx & - k_B \int_{-\infty}^{\infty} \frac{\delta f}{n} \log\left(\frac{f_0}{n}\right) d{\rm v}, \\
\displaystyle
&\approx & - k_B \frac{\Gamma}{n T} \left[ v - \frac{c}{2} + \frac{c^3}{2W} \right],
\edeqa
In comparison to Refs.~\cite{Bizarro_2011_PRE,Bizarro_2015_PoP} where the entropy can decrease in presence of frictions, here we assume an INMDF steady-state without friction. We found that the first order correction of the entropy can either be positive or negative as function of the hidden variables describing the INMDF. We remark that our INMDF is not related to an isolated system, so the inclusion of the collision and the source of energy (e.g., radio-frequency waves, neutral beam, runaway electron) would recover known thermodynamic results of isolated systems (e.g., similar to Ref.~\cite{Bizarro_2015_PoP}). This means that when an external source or sink of energy is turned on, the corrected entropy of the plasma ($\delta s$) computed from an INMDF can increase or decrease with respect to the entropy $s_0$ obtained from a MDF. Both cases are possible without violating the second law of the thermodynamics since here the plasma is non-isolated and the total entropy increase due to the external source is not taken into account. In fact, instead of describing a complete theory of the source and the plasma, we rather focus on the statistical description of the plasma in presence of sources. Then, for a non-isolated system, the maximum value of the entropy is a higher value than $s_0$ as function of the plasma parameters. Moreover, since both variations of $s_I$ are possible, the intermediate case $\delta s=0$ is the solution where the super-thermal particles do not modify globally the entropy of the Maxwellian at the first order. There are $2$ solutions of $\delta s=0$. The first, for finite values of the density and temperature when $\Gamma=0$, recovers the MDF. The second solution is obtained for $c \neq 0$ and $c \neq v$ when
\bgeqa
\label{eq.B.Wcv}
\displaystyle
W &=& \frac{c^2}{1-2\frac{v}{c}}.
\edeqa
The physical interpretation of this specific value of the width of the heat spread still needs to be understood and is still under investigation. However, with this specific INMDF profile, the statistical entropy of the plasma, i.e., without including the entropy increase due to the external sources, is locally constant. This means that the increase of the energy (which increases the entropy) is compensated exactly with the departure of the INMDF with respect to the MDF (which decreases the entropy). The time derivative of the entropy is found by using the Boltzmann equation and omitting the collision operator. It reads $\partial_t s_{(0)} = - {\nabla}  { s}_{(1)}$ where ${ s}_{(k)} = - k_B \int (f/n) \log(f/n) {\rm v}^{k} d{\rm v}$. For the INMDF, $\partial_t s_{I,(0)} = - {\nabla}  { s}_{I,(1)}$ with ${ s}_{I,(0)} = s_I$, and ${ s}_{I,(1)} = { s}_{0,(1)} + \delta { s}_{(1)}$. Using the same expansion of the logarithm, we found that the time evolution of the entropy is modified at the first order by the spatial derivatives of
\bgeqa
\displaystyle
{ s}_{0,(1)} &=& k_B \frac{v}{2} \bigg[ \log(2\pi T) - \frac{v^2}{T} \bigg], \\
\displaystyle
\delta { s}_{(1)} &\approx & k_B \frac{\Gamma}{2 n} \bigg[ \log\left(2\pi T\right) + \frac{v^2}{T} -\frac{2vc}{T} \left( 1-\frac{c^2}{W} \right) \bigg. \nonumber\\
\displaystyle \bigg.
&& \qquad\qquad\qquad\qquad
+ \frac{3 (W+c^2)}{T} - 2 \bigg].
\edeqa
These relations can be used in order to compute the localized statistical entropy evolution of the plasma as a fluid quantity. As a numerical error criteria, for a MDF and without sources, we have to verify that $\partial_t s_{0,(0)} = - {\nabla}  { s}_{0,(1)} = 0$ where $s_{0,(0)} = s_{0}$.\\
The increase or decrease of the entropy in time is given here as function of the spatial gradients of the the hidden variables. A very important interpretation of this result is following:\\
In many experiments (even for other areas than plasmas) non-Maxwellian steady-state distribution functions are observed. This means that in order for us to approximate this steady-state with the proposed INMDF or by creating new analytic NMDFs with as few hidden variables as possible, we have to verify that the local entropy of a non-isolated plasma (including sources) is constant in time because the entropy increase due to the sources is compensated by the entropy reduction of the NMDF steady-state with respect to the maximum entropy of a heated MDF. Then, there cannot be evolving local entropy at NMDF steady-state. From our result of $\partial_t s_{I,(0)} = - {\nabla}  { s}_{I,(1)}$ written as function of the $6$ hidden variables, there are an infinite number of solutions which locally conserve the entropy $\partial_t s_{I,(0)} = 0$ at least at the first order in the expansion of small non-Maxwellian deviations. For example, we obtain one class of these solutions by trivially extracting the density $n$ as function of the other fluid hidden variables $(v,T,\Gamma,c,W)$ since the equation ${ s}_{I,(1)} = { s}_{0,(1)} + \delta { s}_{(1)} = 0$ contains only one occurrence of $n$. We can write this equation as $A+B/n = 0$ with $A={ s}_{0,(1)}$ and $B=n \delta { s}_{(1)}$ which are both functions of $(v,T,\Gamma,c,W)$. This class of solutions of locally constant entropy is obtained when $n = - B/A$, if $A \neq 0$ (i.e., if $v \neq 0$ and $v^2 \neq T \log(2\pi T)$). We have an infinite number of solutions because we have $6$ parameters to resolve the equation $\partial_t s_{I,(0)} = - {\nabla}  { s}_{I,(1)} = 0$. Another class of solution can be found by canceling the derivative of ${ s}_{I,(1)}$. However, even if a decrease of the local entropy can be found in this case, it is an artifact decrease because we intentionally omit the entropy of the external sources which generate the tail (i.e., via the collision operator). This means that by including the collision operator for the self-collisions and the collisions with the sources, the entropy have to be able to increase only following the second law of thermodynamics of isolated systems.\\
As a summary, this result highly suggest the existence of an infinite number of analytic solutions of the Boltzmann equation for non-isolated systems. All these solutions can be obtained as function of the collision frequencies between the plasma and the sources.

The result of the allowed decreasing entropy demystifies the Maxwell's demon since the statistical description of the plasma is still possible even for a non-isolated system. There is no violation of the second law of the thermodynamics which postulates only increasing entropy until the Maxwellian equilibrium of an isolated system. This work could enhance a large range of present technologies since the isolation of a system is very rare. This new approach offers many perspectives since instead of developing more accurate ``interaction'' theories to include very complex external phenomena in a much larger isolated system, we are statistically describing non-isolated systems without dealing with the exact description of the external ``interaction'' theories. Ongoing investigations directly motivated by this result could be on (i) the expansion of the thermodynamics theory for non-isolated systems, (ii) the inclusion of higher orders in $(\delta f/f_0)^k$, (iii) the inclusion of the collision operator in the time evolution of the entropy as function of the hidden variables or, (iv) the numerical observation of NMDF steady states in presence of sources by evolving in time the dynamic equation of the local entropy.

A very large number of other theories usually assume a MDF and can be modified following the three examples shown above. The first goal could be to describe self-consistently the dynamics of dusts, neutrals and impurities from the corrections of measured quantities in presence of non-Maxwellian plasma background. The second goal could be to develop new diagnostics and interpretations of data to systematically measure in experiments non-Maxwellian distribution functions.

\section{Conclusion}
\label{ToC.CCL}
We describe in Sec.~\ref{ToC.NMDFs} the KDF and the sum of MDFs which are successfully used by researchers for numerical post-analyses. The KDF observed in astrophysics is one example of NMDF, even if to date there is no physical interpretation of the parameter $\kappa$. Another choice of NMDF commonly used is the bi-modal distribution function (the sum of two MDFs) but this choice can be consistent only at a very specific collisionality regime (the self-collisionality of the bulk and the fast population are both much larger than the collisionality of the interaction between them). Moreover, a new function called the INMDF is introduced here in order to offer more choices (consistent with a finite collisionality) for the representation of NMDFs. Readers can create as new analytic distribution functions as needed to represent the phenomenon under investigation in such a way that the velocity phase space integrals are analytically manageable as function of as few hidden variables as possible. In comparison to the bi-modal distribution function, the INMDF is a possible solution for other collisionality regimes since it describes a displacement of a population of particles from one energy to another. As a summary of this work, INMDFs are new proposed functions that seem much more efficient in describing non-thermalized plasmas than any existing formula thanks to the low number of required parameters (i.e., hidden variables). Moreover, INMDFs help to understand unsolved problems such as the ones at the origin of non-Maxwellian bulks observed in JET and TFTR, thanks to the physical interpretation of the hidden variables. With the distribution functions described in Sec.~\ref{ToC.NMDFs}, it is shown in Sec.~\ref{ToC.KinCorrections} that analytic predictions of kinetic corrections are possible and examples on the SEE, the Langmuir probe characteristic curve and the entropy are shown. Because these examples are not directly related to a specific plasma, {this proves the universal property of the INMDFs and the physical interest to consider non-orthogonal basis sets.} More applications will be found later. Moreover, other details of NMDFs will be published elsewhere such as the fluid reduction (see Ref.~\cite{Izacard_2016_Paper2}), which opens the access to the next generation of fluid codes by including non-collisional and collisional kinetic effects (see Ref.~\cite{Izacard_2016_FischSymposium}), or the description of the transport. Indeed, some results suggest that fluid models can be similar to kinetic codes such as the observation of asymmetric heat flux inside an island by using a fluid model with some finite Larmor radius terms~\cite{Izacard_2016_PoP}, using profiles of transport coefficients~\cite{Chankin_2007_NF_47a,
Chankin_2007_NF_47b,
Groth_2013_NF_53} to reduce the radiation shortfall or using nonlocal fluid closures~\cite{Held_2004_PoP,Dimits_2013_PoP}. A clarification of these examples will be detailed later. \\
For the physical interpretation of the kinetic effects shown in Sec.~\ref{ToC.NMDFs.Reality} and Sec.~\ref{ToC.KinCorrections}, we found four groundbreaking results
\begin{itemize}
\item[(i)] The physical reality of the INMDFs introduced here is proved by the over plot fitting of the numerical NMDF model obtained in Ref.~\cite{Beausang_2011_RSI} for the explanation of the TS-ECE discrepancy of the electron temperature interpretation in JET. We found that some particles at low energy are heated and others at higher energy are cooled down. This interpretation would help to understand the physical origin of the NMDF observed in JET and to resolve discrepancies between diagnostics due to the presence of NMDFs.
\item[(ii)] The presence of super-thermal particles in the Langmuir probe characteristic curve induces diffusion effects which have been commonly observed by using a diffusion term in the formula when a MDF was assumed. This result highly suggests more investigations to replace the usual ad-hoc dissipative coefficients and the diffusion terms by NMDFs.
\item[(iii)] The use of the analytic computation of the SEE in presence of NMDFs allows us the unexpected observation that the empirical formula of the SEE published in Ref.~\cite{Bacharis_2010_PoP_17} is not consistent with a MDF but is consistent with a presence of $\sim 3\%$ of super-thermal particles. Moreover, because the sum of $2$ MDFs {does not} reproduce much better the empirical formula {(in contrary to the INMDF)}, it suggests that the collisionality {in Ref.~\cite{Bacharis_2010_PoP_17}} was {not} negligible or infinite, but finite as explained above.
\item[(iv)] Finally, the physical motivation of the INMDF given by \refeq{eq.nM.fINMDF} is shown by the explicit decrease of the entropy for a non-isolated system without violating the second law of thermodynamics. This is the first simple analytic function which can be consistent with NMDFs at finite collisionality in presence of external source of energy. With this work, the entropy is not viewed as a degree of disorder but is interpreted as a level of sharing information (or energy) between particles of the plasma.
\end{itemize}
From these results obtained thanks to our new INMDFs, many perspectives are possibles such as the description of ion orbit losses~\cite{Stacey_2013,Stacey_2016} and the fast ions~\cite{Fasoli,Fasoli_NaturePhys}, or the development of experimental measurements of NMDFs using for example Thomson scattering~\cite{Luna_2003_RSI,Beausang_2011_RSI} or Langmuir probes. In summary, because different measurements can be more or less sensitive to the presence of super-thermal particles, better interpretations of experimental data are possible and more accurate measurement techniques of the distribution function can be developed. 

\section*{Acknowledgements}
\addcontentsline{toc}{section}{Acknowledgements}
This work was supported by the LLNL Postdoctoral independent research funding. The author would like to acknowledge M.~Jaworski (PPPL) for our discussions on Langmuir probes measurements, and B.~Cohen (LLNL) and D.P.~Brennan (Princeton University) for their very helpful comments on this manuscript
. This work was performed under the auspices of the U.S. Department of Energy by Lawrence Livermore National Laboratory under Contract DE-AC52-07NA27344.

\appendix
\section*{Appendix}
\addcontentsline{toc}{section}{Appendix}
In these appendixes, we detail analytic computations obtained from Mathematica and Ref.~\cite{GR_2007}. {The reduction from special functions such as the hypergemoetric $ _1F_1(a,b,z)$ or the incomplete $\Gamma(x,z)$ functions to a small number of terms is possible because of the evaluation of these functions at specific values. All velocity phase space integrals of any INMDF can be analytically computed following these appendices.}

\section{Integrals of Kappa distributions}
\label{ToC.App.KqkappaTa}
From the definition of $K_q(\kappa,T)$ given by \refeq{eq.nM.KqkappaT} 
we found
\bgeqa
\displaystyle
K_{q}(\kappa ,T) &=& \frac{1+(-1)^q}{2} W_{\kappa}^{(q+1)/2} \nonumber\\
\displaystyle
&&\qquad\quad \times
\frac{ \Gamma\left(\frac{1+q}{2}\right) \Gamma\left(\frac{1-q}{2}+\kappa\right) }{ \Gamma(1+\kappa) }.
\edeqa
From the definition of $K_q(\kappa,T,a)$ given by \refeq{eq.nM.KqkappaTa} 
we found
\bgeqa
\displaystyle
K_{q}(\kappa ,T,a) &=&  \frac{W_{\kappa}^{(\kappa+1)}}{(1-q+2\kappa) a^{(1-q+2\kappa)}} \nonumber\\
\displaystyle
&& \quad \times
_2F_1\left(\kappa +1,\kappa_1 ;\kappa_2 ;-\frac{W_{\kappa}}{a^2}\right),
\edeqa
with $a \neq 0$, $\kappa_1=\kappa-(q-1)/2$ and  $\kappa_2=\kappa-(q-3)/2$, where $_2F_1$ is the hypergeometric function and
\bgeqa
\displaystyle
K_{q}(\kappa ,T,0) &=& W_{\kappa}^{(q+1)/2} \frac{ \Gamma\left(\frac{1+q}{2}\right) \Gamma\left(\frac{1-q}{2}+\kappa\right) }{ 2 \Gamma(1+\kappa) }.
\edeqa

\section{Integrals of Maxwellian distributions}
\label{ToC.App.Jkab}
From the definition of $J_k(a,b)$ given by \refeq{eq.Gen.Jkab} 
we found
\bgeqa
\displaystyle
J_k(a,b) &=& \frac{a^{-(\frac{k}{2}+1)}}{2}  \bigg[ \left(1 - (-1)^k\right) b\ \Gamma\left(\frac{k}{2}+1\right)\ \Bigg. \nonumber\\
\displaystyle
\Bigg. && \qquad\qquad
\times \ _1F_1\left(\frac{k}{2}+1,\frac{3}{2},\frac{b^2}{4a}\right) \bigg. \nonumber\\
\displaystyle\bigg.
&&+ \left((-1)^k+1\right) \sqrt{a}\ \Gamma\left(\frac{k+1}{2}\right)\ \Bigg. \nonumber\\
\displaystyle
\Bigg. && \qquad\qquad
\times \ _1F_1\left(\frac{k+1}{2},\frac{1}{2},\frac{b^2}{4a}\right) \bigg],
\edeqa
with $k \in \mathbb{N}$ and the Kummer confluent hypergeometric function of the first kind $_1F_1$
\bgeqa
\displaystyle
_1F_1(p,q,r) = \sum_{n=0}^{\infty} \frac{(p)_n}{(q)_n} \frac{r^n}{n!},
\edeqa
where $(p)_0=1$ and $(p)_n = \frac{(p+n-1)!}{(p-1)!}$. However, this definition of the Kummer confluent hypergeometric function with the sum of an infinite number of term turns out to have a finite number of terms for the values $(p,q)$ of interest (i.e., $p \in \{ 3/2, 2, 5/2, 3, 7/2 \}$ and $q \in \{ 1/2, 3/2 \}$) as
\bgeqa
\displaystyle
_1F_1\left(1,\frac{1}{2},z\right) &=& 1 + Z, \\
\displaystyle
_1F_1\left(1,\frac{3}{2},z\right) &=& \frac{1}{2z} Z, 
\edeqa
\bgeqa
\displaystyle
_1F_1\left(2,\frac{1}{2},z\right) &=& 1 + z + \left( \frac{3}{2}  + z \right) Z, \\
\displaystyle
_1F_1\left(2,\frac{3}{2},z\right) &=& \frac{1}{2} + \frac{1}{2} \left( 1 + \frac{1}{2z} \right) Z, 
\edeqa
\bgeqa
\displaystyle
_1F_1\left(3,\frac{1}{2},z\right) &=& 1 + \frac{9}{4} z + \frac{1}{2} z^2 \nonumber\\
\displaystyle
&&\qquad
+ \frac{1}{2} \left( \frac{15}{4} + 5 z + z^2 \right) Z, \\
\displaystyle
_1F_1\left(3,\frac{3}{2},z\right) &=& \frac{5}{8} + \frac{1}{4} z + \frac{1}{4} \left( \frac{3}{4 z} + 3 + z \right) Z,\ \ 
\edeqa
\bgeqa
\displaystyle
_1F_1\left(\frac{1}{2},\frac{1}{2},z\right) &=& \exp(z), \\
\displaystyle
_1F_1\left(\frac{3}{2},\frac{1}{2},z\right) &=& (1 + 2 z) \exp(z), \\
\displaystyle
_1F_1\left(\frac{3}{2},\frac{3}{2},z\right) &=& \exp(z), 
\edeqa
\bgeqa
\displaystyle
_1F_1\left(\frac{5}{2},\frac{1}{2},z\right) &=& \left( 1 + 4 z + \frac{4}{3} z^2 \right) \exp(z), \\
\displaystyle
_1F_1\left(\frac{5}{2},\frac{3}{2},z\right) &=& \left( 1 + \frac{2}{3} z \right) \exp(z), \\
\displaystyle
_1F_1\left(\frac{7}{2},\frac{1}{2},z\right) &=& \left( 1 + 6 z + 4 z^2 + \frac{8}{15} z^3 \right) \nonumber\\
\displaystyle
&&\qquad\qquad\qquad\quad \times \exp(z), \\
\displaystyle
_1F_1\left(\frac{7}{2},\frac{3}{2},z\right) &=& \left( 1 + \frac{4}{3} z + \frac{4}{15} z^2 \right) \exp(z),
\edeqa
where $Z = \sqrt{\pi} \sqrt{z} \exp(z) {\rm Erf}\left(\sqrt{z}\right)$. Then the first terms of $J_k(a,b)$ are
\bgeqa
\displaystyle
J_{0}(a,b) &=& a^{-1/2} C, \\
\displaystyle
J_{1}(a,b) &=& \frac{a^{-3/2}}{2} b C, \\
\displaystyle
J_{2}(a,b) &=& \frac{a^{-5/2}}{4} (2a+b^2) C, \\
\displaystyle
J_{3}(a,b) &=& \frac{a^{-7/2}}{8} (6a+b^2) b C, \\
\displaystyle
J_{4}(a,b) &=& \frac{a^{-9/2}}{16} (12a^2+12ab^2+b^4) C, \\
\displaystyle
J_{5}(a,b) &=& \frac{a^{-11/2}}{32} (60a^2+20ab^2+b^4) b C, \\
\displaystyle
J_{6}(a,b) &=& \frac{a^{-13/2}}{64} (120a^3+180a^2b^2 \nonumber\\
\displaystyle
&&\qquad\qquad\quad +30ab^4+b^6) C,
\edeqa
and so forth and so on, with $C = \sqrt{\pi} \exp\left(\frac{b^2}{4a}\right)$ if $a \in \mathbb{R}^{+\star}$.

\section{Useful relations for the secondary electron emission}
\label{ToC.App.Jkab0}
From the definition of $J_k(a,b,0)$ given by \refeq{eq.SEE.Jkab0} 
we found a generalization of the function $J_k(a,b,0)$
\bgeqa
\displaystyle
&&J_k(a,b,0) = a^{-(k+2)}  \bigg[ a \Gamma(k+1)\ _1F_1\left(k+1,\frac{1}{2},\frac{b^2}{4a}\right) \bigg. \nonumber\\
\displaystyle\bigg.
&&\qquad\ + b\sqrt{a}\ \Gamma\left(k+\frac{3}{2}\right)\ _1F_1\left(k+\frac{3}{2},\frac{3}{2},\frac{b^2}{4a}\right) \bigg],
\edeqa
with the Kummer confluent hypergeometric functions of the first kind $_1F_1$ given in App.~\ref{ToC.App.Jkab}. Then the first terms of $J_k(a,b)$ are
\bgeqa
\displaystyle
J_{0}(a,b,0) &=& \frac{a^{-1/2}}{2} C, \\
\displaystyle
J_{1}(a,b,0) &=& \frac{a^{-3/2}}{4} \left[   b C  +  2\sqrt{a}   \right], \\
\displaystyle
J_{2}(a,b,0) &=& \frac{a^{-5/2}}{8} \left[   (2a+b^2) C  +  2b\sqrt{a}   \right], \\
\displaystyle
J_{3}(a,b,0) &=& \frac{a^{-7/2}}{16} \big[   b (6a+b^2) C  \big.\nonumber\\
\displaystyle
\big. &&\qquad\quad +  2\sqrt{a}(4a+b^2)   \big], \\
\displaystyle
J_{4}(a,b,0) &=& \frac{a^{-9/2}}{32} \big[   (12a^2+12ab^2+b^4) C \big. \nonumber\\
\displaystyle
\big. && \qquad\quad + 2b\sqrt{a}(10a+b^2)   \big], \\
\displaystyle
J_{5}(a,b,0) &=& \frac{a^{-11/2}}{64} \big[   b(60a^2+20ab^2+b^4) C  \big. \nonumber\\
\displaystyle
\big. &&\qquad\quad + 2\sqrt{a}(2a+b^2)(16a+b^2)   \big],\quad
\edeqa
with $C = \sqrt{\pi} \exp\left(\frac{b^2}{4a}\right) \left( 1 + {\rm Erf}\left(\frac{b}{2\sqrt{a}}\right)\right)$ and ${\rm Erf}(z)$ is the error function at $z$.

\section{Useful relations for the Langmuir probe interpretation}
\label{ToC.App.Jkabc}
From the definition of $J_k(a,b,c)$ given by \refeq{eq.LP.Jkabc} 
and by using the change of variable $x={\rm v}-b/(2a)$ we found
\bgeqa
\displaystyle
J_k(a,b,c) &=& \exp\left(\frac{b^2}{4a}\right) \int_{C}^{\infty} \left( x+\frac{b}{2a} \right)^k \exp\left(-ax^2\right) dx, \nonumber\\
\displaystyle
&=& \exp\left(\frac{b^2}{4a}\right) \sum_{n=0}^k \left(\begin{array}{c}k\\n\end{array}\right) \left(\frac{b}{2a}\right)^{k-n} \nonumber\\
\displaystyle
&&\qquad\qquad\qquad\qquad \times
\frac{\Gamma\left(\frac{n+1}{2},aC^2\right)}{2a^{(n+1)/2}},
\edeqa
with $C=c-b/(2a)$, the help of the equation 3.381.9 on page 346 of \cite{GR_2007} where $\Gamma(a,b)$ is the incomplete Euler Gamma function detailed in App.~\ref{ToC.App.Gamma}. The first terms of $J_k(a,b,c)$ are
\bgeqa
\displaystyle
J_0(a,b,c) &=& \mathcal{G}_2, \\
\displaystyle
J_1(a,b,c) &=&  \frac{1}{2a} \left[ \mathcal{E}_2 + b \mathcal{G}_2 \right], \\
\displaystyle
J_2(a,b,c) &=& \frac{1}{(2a)^2} \left[ (b+2ac) \mathcal{E}_2 + (b^2+2a) \mathcal{G}_2 \right], \\
\displaystyle
\displaystyle
J_3(a,b,c) &=&  \frac{1}{(2a)^3} \big( b^2+4a+2ac(b+2ac) \big) \mathcal{E}_2 \nonumber\\
\displaystyle
&&
+ \frac{1}{(2a)^3} b (b^2+6a) \frac{\mathcal{G}_2}{\mathcal{E}_2}, \\
\displaystyle
J_4(a,b,c) &=& \frac{\mathcal{E}_2}{(2a)^4} \big( (b+2ac) (b^2+4a^2c^2) + 12a^2c + 10ab \big) \nonumber\\
\displaystyle
&&
+ \frac{1}{(2a)^4} (12a^2 + 12ab^2 + b^4) \frac{\mathcal{G}_3}{\mathcal{E}_2}, 
\edeqa
with the following definitions
\bgeqa
\displaystyle
\mathcal{E}_1 &=& \exp\left( \frac{b^2}{4a} \right), \\
\displaystyle
\mathcal{E}_2 &=& \exp\left( -ac^2+bc \right), \\
\displaystyle
\mathcal{G}_1 &=& {\rm Erf}\left(\frac{b-2ac}{2\sqrt{a}} \right), \\
\displaystyle
\mathcal{G}_2 &=& \frac{1}{2} \sqrt{\frac{\pi}{a}}  \mathcal{E}_1 \left( 1 + \mathcal{G}_1 \right), \\
\displaystyle
\mathcal{G}_3 &=& \frac{1}{2} \sqrt{\frac{\pi}{a}}  \mathcal{E}_1 \left( 1 + (b-2ac) \mathcal{G}_1 \right). 
\edeqa

\section{Useful relations for the incomplete Gamma function}
\label{ToC.App.Gamma}
The used form of the incomplete Gamma function are
\bgeqa
\displaystyle
\Gamma\left(n+1,x\right) &=& n!\ \exp(-x) \sum_{k=0}^{n} \frac{x^k}{k!},\\
\displaystyle
\Gamma\left(n+\frac{1}{2},x\right) &=& \Gamma\left(n+\frac{1}{2}\right){\rm Erfc}\left(\sqrt{x}\right) \nonumber\\
\displaystyle
&+& (-1)^{n-1} \exp(-x) \sqrt{x} \nonumber\\
\displaystyle
&\times &
\sum_{k=0}^{n-1} \left(\frac{1}{2}-n\right)_{(n-1-k)} (-x)^k,
\edeqa
with $n \in \mathbb{N}^{\star}$, $\Gamma(0)=1$, $\Gamma(k,0)=\Gamma(k)$, $(u)_k$ is the Pochhammer symbol and the first terms of the incomplete Gamma function read
\bgeqa
\displaystyle
\Gamma(1,x) &=& \exp\left(-x\right), \\
\displaystyle
\Gamma(2,x) &=& \left(1+x\right) \exp\left(-x\right), \\
\displaystyle
\Gamma\left(\frac{1}{2},x\right) &=& X, \\
\displaystyle
\Gamma\left(\frac{3}{2},x\right) &=& C + \frac{1}{2} X, \\
\displaystyle
\Gamma\left(\frac{5}{2},x\right) &=& \left(x+\frac{3}{2}\right) C + \frac{3}{4} X, 
\edeqa
where $C=\sqrt{x} \exp(-x)$, $X=\sqrt{\pi} {\rm Erfc}\left(\sqrt{x}\right)$ and ${\rm Erfc}(x)=1-{\rm Erf}(x)$ is the inverse error function.

\section{Entropy corrections}
\label{ToC.App.Entropy}
From the entropy defined by \refeq{eq.B.s} 
we found using the previously defined function $J_k(a,b)$ and with the fact that
\bgeqa
\displaystyle
\log\left(\frac{f_0}{n}\right) &=& \log\left(\frac{\Delta}{n}\right) - \frac{1}{2T} {\rm v}^2 + \frac{v}{T} {\rm v},
\edeqa
that the entropy for a MDF is
\bgeqa
\displaystyle
s_0 &=& - k_B \int_{-\infty}^{\infty} \frac{\Delta}{n} \exp\left(- \frac{1}{2T} {\rm v}^2 + \frac{v}{T} {\rm v}\right) \nonumber\\
\displaystyle 
&&\quad \times
\bigg[ \log\left(\frac{\Delta}{n}\right) - \frac{1}{2T} {\rm v}^2 + \frac{v}{2T} {\rm v} \bigg] d{\rm v},
\\
\displaystyle
&=& - k_B \frac{\Delta}{n} \bigg[ \log\left(\frac{\Delta}{n}\right) J_0\left( \frac{1}{2T},\frac{v}{T}\right) \bigg. \nonumber\\
\displaystyle \bigg.
&&- \frac{1}{2T} J_2\left( \frac{1}{2T},\frac{v}{T}\right) + \frac{v}{T} J_1\left( \frac{1}{2T},\frac{v}{T}\right) \bigg],
\edeqa
\bgeqa
\displaystyle
s_0 
&=& - k_B \frac{\Delta}{n} \bigg[ \log\left(\frac{\Delta}{n}\right) \sqrt{2\pi T} \exp\left(\frac{v^2}{2T}\right) \bigg. \nonumber\\
\displaystyle \bigg.
&&\qquad - \frac{1}{2T} \sqrt{2\pi T} \exp\left(\frac{v^2}{2T}\right) (T+v^2) \nonumber\\
\displaystyle \bigg.
&&\qquad + \frac{v}{T} \sqrt{2\pi T} \exp\left(\frac{v^2}{2T}\right) v \bigg],
\edeqa
\bgeqa
\displaystyle
s_0 &=& - k_B \bigg[ \log\left(\frac{\Delta}{n}\right) - \left(\frac{1}{2}+\frac{v^2}{2T}\right) + \frac{v^2}{T} \bigg],
\edeqa
\bgeqa
\displaystyle
s_0 &=& \frac{k_B}{2} \big[ 1 + \log(2\pi T) \big],
\edeqa
Then,
\bgeqa
\displaystyle
s_I &=& - k_B \int_{-\infty}^{\infty} \left(\frac{f_0}{n} + \frac{\delta f}{n}\right) \Bigg( \log\left( \frac{f_0}{n} \right) \Bigg. \nonumber\\
\displaystyle
\Bigg. && \qquad\qquad\qquad\qquad\qquad + \log\left( 1+\frac{\delta f}{f_0} \right) \Bigg) d{\rm v}, \nonumber\\
\displaystyle
&=& s_0 - k_B \int_{-\infty}^{\infty} \frac{\delta f}{n} \log\left( \frac{f_0}{n} \right) d{\rm v} \nonumber\\
\displaystyle
&+& k_B \sum_{k=1}^{\infty} \frac{(-1)^k}{k} \int_{-\infty}^{\infty} \left(\frac{f_0}{n} + \frac{\delta f}{n}\right) \left( \frac{\delta f}{f_0} \right)^{k} d{\rm v}.\quad
\edeqa
\bgeqa
\displaystyle
&&s_I = s_0 - k_B \int_{-\infty}^{\infty} \frac{\delta f}{n} \log\left( \frac{f_0}{n} \right) d{\rm v} \nonumber\\
\displaystyle
&&+ \frac{k_B}{n} \sum_{k=1}^{\infty} \frac{(-1)^k}{k} \int_{-\infty}^{\infty} \left. \frac{\left( \delta f \right)^{k}}{\left(f_0\right)^{k-1}} + \frac{\left( \delta f \right)^{k+1}}{\left(f_0\right)^{k}} \right. d{\rm v}.\ 
\edeqa
The first order ($k=1$) of the last integral is proportional to $\int_{-\infty}^{\infty} \delta f d{\rm v} = 0$. This last sum can be written with the fact that
\bgeqa
\displaystyle
\left( f_0 \right)^k &=& \Delta^k \exp\left(- \frac{k}{2T} {\rm v}^2 + k \frac{v}{T} {\rm v} \right), \\
\displaystyle
\left( \delta f \right)^k &=& \Delta_I^k ({\rm v}-c)^k \exp\left(- \frac{k}{2W} {\rm v}^2 + k \frac{c}{W} {\rm v} \right),\quad
\edeqa
and by using the previously defined function $J_k(a,b)$, as
\bgeqa
\displaystyle
&&\sum_{k=1}^{\infty} \frac{(-1)^k}{k(k+1)} \Delta \left( \frac{\Delta_I}{\Delta} \right)^{k+1} \sum_{m=0}^{k+1} \left(\begin{array}{c}k+1\\ m\end{array}\right) (-c)^{k+1-m} \nonumber\\
\displaystyle
&&\qquad \times J_m\left( \frac{k+1}{2W}-\frac{k}{2T}, \frac{(k+1)c}{W}-\frac{kv}{T} \right).
\edeqa
Then $s_I = s_0 + \delta s$ with
\bgeqa
\displaystyle
&&\delta s = - k_B \frac{1}{n} \int_{-\infty}^{\infty} \delta f \log\left( \frac{f_0}{n} \right) d{\rm v} \nonumber\\
\displaystyle
&+& k_B \frac{\Delta}{n} \sum_{k=1}^{\infty} \frac{(-1)^k}{k(k+1)} \left( \frac{\Delta_I}{\Delta} \right)^{k+1} \sum_{m=0}^{k+1} (-c)^{k+1-m} \nonumber\\
\displaystyle
&\times & \left(\begin{array}{c}k+1\\ m\end{array}\right) J_m\left( \frac{k+1}{2W}-\frac{k}{2T}, \frac{(k+1)c}{W}-\frac{kv}{T} \right). \qquad\ 
\edeqa

\addcontentsline{toc}{section}{References}
\begin{flushleft}
\begin{footnotesize}

\end{footnotesize}
\end{flushleft}

\end{document}